\documentclass{jfm}
\usepackage{graphicx}
\usepackage{epstopdf,epsfig}
\usepackage{newtxtext}
\usepackage{newtxmath}
\usepackage{natbib}
 \usepackage{multirow} 
\usepackage{hyperref}
\usepackage{booktabs}
\usepackage[dvipsnames]{xcolor}
\definecolor{myblue}{RGB}{0, 0, 230}
\hypersetup{
    colorlinks = true,
    linkcolor = myblue,
    urlcolor   = myblue,
    citecolor  = myblue,
}

\newcommand{\RomanNumeralCaps}[1]
\linenumbers

\usepackage{lineno}

\shorttitle{Shock propagation through a local constriction}
\shortauthor{R. Heppner, H. Chandravamsi, Y.Gichon, S. H. Frankel,  and O. Ram}

\title{Shock propagation through a local constriction}

\author{
Raz Heppner,\aff{1}\aff{2}
Hemanth Chandravamsi,\aff{1}\aff{3}
Yoav Gichon,\aff{1}
Steven H. Frankel \aff{1}
\and
Omri Ram\aff{1}\corresp{\email{omri.ram@technion.ac.il}}
}

\affiliation{
\aff{1} Faculty of Mechanical Engineering, Technion - Israel Institute of Technology,
Haifa 3200003, Israel

\aff{2} Grand Technion Energy Program, Technion - Israel Institute of Technology,
Haifa 3200003, Israel

\aff{3} Department of Mechanical Engineering, Johns Hopkins University, Baltimore, MD 21218, USA
}
\begin{document}

\date{\today}

\maketitle 

\begin{abstract}
The interaction of a shock wave with a localized constriction in a straight conduit is investigated by systematically varying the blockage ratio in the range 0.35-0.75, the normalized constriction length in the range 0.25-2, and the incident Mach numbers of 1.4 and 1.8. Abrupt rectangular constrictions and smoothly contoured sinusoidal constrictions are considered, as they provide two limiting configurations. Validated Large-eddy simulations resolve both the transient start-up dynamics and the subsequent propagation of reflected and transmitted shock waves. The results show that, for rectangular constrictions, the reflected shock strength depends primarily on the blockage ratio and is largely independent of length, whereas the transmitted shock exhibits measurable sensitivity to constriction length. In contrast, sinusoidal constrictions display a strong coupling between blockage and length, with the reflection process governed by the local contour slope and evolving reflection topology. The start-up process within the constriction occurs over time scales one to two orders of magnitude longer than the shock passage time and is characterized by a sequence of reflection, separation, and flow reorganization events that determine the eventual steady shock configuration. At later times, the reflected shock Mach number scales linearly with blockage ratio, while the transmitted shock strength decreases monotonically with increasing blockage. Based on these trends, semi-empirical models are developed to predict the strengths of both reflected and transmitted shocks across the parameter space considered. These results provide a unified framework for understanding and predicting shock propagation in conduits with localized geometric variations, with direct relevance to compressible internal flows in engineering and natural systems.
\end{abstract}

\section{Introduction}

Moving shock waves often form in industrial processes by accident or by design in cases that involve, for example, changes in combustion regimes, runaway chemical reactions, or the release of high-pressure gas or combustion products \citep{golub2007shock, duan2015experimental,cheng2019full, gong2023numerical}. These could propagate in pipes and through various systems, causing impulsive loading that can damage valves and measuring devices. Eventually, the shock might discharge the system, propagate in open space, and become a hazard to personnel in close range or cause noise disturbances at longer ranges. Hence, industrial facilities often install automatic pressure valves, passive buffers, and mufflers to mitigate these risks \citep{canua2025review,ChristianJenni2023}. Transient shock wave propagation also occurs in high-speed propulsion systems, primarily during the startup processes of supersonic engine intakes, scramjet engines, and rockets \citep{arad2024simulation}. These, typically strong shock waves, interact with the complex internal geometry of the engine, causing non-uniform loading and vibration, which might lead to immediate catastrophic damage or shorten the lifetime of the components due to localized straining. Accidental or intentional explosions can occur in underground tunnels, including the entrances to shelters, long transportation tunnels, and mines \citep{wang2021experimental, lin2022study,gan2024blast}. In addition, understanding shock propagation in nonuniform conduits is important for modeling and predicting natural processes, for example, the dynamics of explosive volcanic eruptions, in which gas jets pass through constricted vents and craters \citep{ogden2011fluid, medici2014modeling}. In these cases, internal explosions will lead to the propagation of strong shock waves in the tunnel, which typically has a non-uniform internal geometry, including structural reinforcements, transitions between sections of different cross-sectional areas, vents, splits, and turns. 

As a shock wave propagates in a conduit and encounters a local obstruction or constriction, it interacts with the geometry, triggering a complex, transient series of events. These include multiple shock-walls, shock-shock, and shock-flow interactions. The exact morphology and subsequent evolution of these reflections, as well as the developing flow field, depend heavily on the specific obstruction geometry. The interaction of the shock wave with the obstruction initiates a transient start-up process in which the impulsively formed flow behind the incident shock wave conforms to the contour of the constriction, and the shock wave reflections subside, reaching a steady state. Depending on the blockage ratio and the specific geometry of the narrow region, a strong shock is reflected back upstream and propagates against the incoming flow. Since a portion of the energy has been invested in the formation of the reflected shock wave, the incident shock that propagates downstream of the constriction will be weaker. As one might expect, smoother, more gradually contoured geometries create weaker reflected shock waves upstream and stronger transmitted shock downstream, and vice versa \citep{sloan1975model, igra2001experimental, gaetani2008shock, berger2010experimental}. 

Numerous studies have examined the interaction between shock waves and localized geometries, including abrupt area changes, porous barriers, and internal constrictions. Earlier works such as those by \citet{monroe1959investigation} and \citet{dadone1971interaction} examined the reflection and transmission of shock waves through orifice plates embedded in ducts, establishing early models for estimating transmitted shock strength and the onset of choking. Later efforts advanced these ideas by exploring more complicated geometries. \citet{berger2010experimental, berger2015experimental} conducted systematic experimental studies of both single and multiple constrictions in a shock tube, showing that the constriction shape, blockage ratio, and configuration influence the attenuation of shock loads and the resulting wave patterns. These results emphasized that, at high blockage ratios of approximately 0.375, 0.5, and 0.625, geometry-induced flow features like vortex generation and shock trapping significantly affect the pressures recorded at downstream surfaces. These experimental observations have been supported by numerical studies, such as those by \citet{rajaseker2023numerical}, who modeled the interaction of blast waves with barriers of varying shape and spacing, and \citet{rigby2023blast}, who summarized the effects of various types of constrictions on blast wave loadings in urban scenarios.
The propagation of a shock wave in a conduit of varying geometry and its influence on the shock has been explored through both theory and experiments for both sharp area change \citep{gottlieb1983interaction, gichon2024dynamics,kickliter2025numerical}  and with varying geometries \citep{bird1959effect,igra1994shock, igra1998nonstationary,weiss2010behavior, dowse2014area,hermet2021shock}. These studies showed that smoother contractions led to more gradual shock adaptation and reduced reflection intensity compared to sharp, step-wise area changes, which generated stronger reflected shocks and more pronounced flow separation near the contraction. These studies highlighted that the geometry of the area reduction, not just the contraction ratio, plays a crucial role in determining the strength of downstream transmitted shocks and the onset of complex flow features, such as boundary-layer detachment and vortex generation. These findings underscore the importance of considering both the curvature and the smoothness of the constriction when modeling shock-geometry interactions, as they directly affect the effectiveness of energy transmission through confined flows. In addition, they showed that a long smooth contraction significantly reduced the amplitude and duration of downstream pressure fluctuations by promoting more gradual flow adaptation.\\
\indent Few studies have directly examined the effects of local constriction of the cross-section where the area construction is followed by expansion to the same cross-section. Some information can be extracted from studies on the effects of shock interaction with Porous and perforated barriers, which have been widely studied as a means to attenuate shock waves by modulating the transmitted flow through arrays of small openings. \citet{britan2006shock} modeled shock interaction with orifice plates and demonstrated that the plate thickness, porosity, and internal geometry of the holes significantly influence the transmitted shock strength and the generation of secondary wave structures. \citet{ram2018pressure} showed that multiple perforated plates can create resonant cavities where shock reflections and reverberations are governed not only by porosity but also by the spacing between plates and the type of gas. \citet{seeraj2009dual, majji2024predicting,majji2025attenuation} further found that directional porosity, achieved through staggered, angled holes, enhances impulse mitigation by dispersing the transmitted front into multiple weaker waves. While these studies focus on extended porous barriers, the physical principles underlying attenuation, namely, rapid area changes, flow choking, and shock diffraction through confined openings, are closely related to those present in localized geometric constriction. Recent experimental studies \citep{langdon2011influence, schunck2021blast} and numerical simulations \citep{kumar2017attenuation} also highlight the analogy that when a shock interacts with a porous medium, each pore behaves as a localized nozzle or contraction. The resulting flow features, such as shock focusing, vortex shedding, and boundary layer separation within the pore, mirror the interactions observed when a shock passes through a short constriction in a duct. In all cases, the blockage ratio, length scale, and internal curvature of the flow path play critical roles in determining the ratio between transmitted and reflected shock strength and in governing the downstream impulsive flow field. Thus, although porous barriers and duct contractions differ in global geometry, their underlying physics is fundamentally similar, making studies of one highly relevant to understanding the other.\\
\indent Taken together, existing studies provide a solid foundation for understanding how blockage ratio and contraction shape influence the multiscale shock-wall, shock-shock, and shock-flow interactions that arise when a moving shock propagates through a constriction section of a straight conduit. These works have established key trends in reflected and transmitted shock strengths and have clarified the role of abrupt versus gradual area changes in shaping the global wave dynamics. Nevertheless, several aspects of the problem remain only partially resolved, motivating the present study. In particular, most available investigations explore a restricted subset of geometric parameters, typically varying either the blockage ratio, the contraction length, or an overall area change while holding other degrees of freedom fixed. As a result, the coupled influence of these parameters on the redistribution of energy between reflected and transmitted waves, as well as on the structure of the impulsively generated downstream flow, has not been systematically characterized. A unified view of how blockage and length act together to shape the transient shock system is therefore still lacking. Moreover, prior work has largely focused on canonical geometries such as sharp orifices, steps, or single, idealized nozzle contours. While these configurations have yielded valuable physical insight, they do not span the broader class of smoothly contoured constrictions in which curvature and surface smoothness can significantly alter the transient build-up of the reflected shock system, the approach to choking, and the emergence of jetting and secondary shock structures. Finally, the start-up process itself, defined as the short-time, highly unsteady adjustment from the passage of an incident planar shock to a quasi-established flow configuration, has received comparatively limited attention. Open questions remain regarding the temporal sequence, sensitivity, and geometric dependence of key flow features, including separation regions, shear-layer formation, and the establishment of standing shocks within the throat. Addressing these gaps requires experiments and analysis that resolve both the geometric parameter space and the early-time transient dynamics, which is the focus of the present work.

In this study, we address this gap by systematically varying both the blockage ratio and the axial length of a smooth, localized constriction, comparing it to an abrupt step constriction. We resolve the startup dynamics using high-order implicit LES, which are validated with time-resolved schlieren and wall-pressure measurements. We (i) quantify how the coupled \(h/H,\,L/H\) parameters control the rate at which the reflected system develops and the ultimate balance between reflected and transmitted shocks; (ii) identify geometry-dependent pathways for secondary structure formation of standing shocks, expansion fans, shear-layer separation and jetting in the throat and their signatures in pressure histories; and (iii) propose reduced order semi-empirical models for predicting the late-time transmitted and reflected shock waves strengths.

\section{Problem description}

\begin{figure}
   \centerline{\includegraphics[scale=0.75]{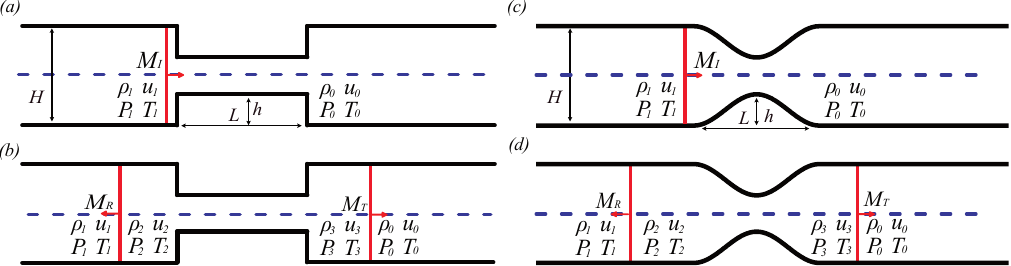}}
   \caption{Schematic figure of the problem, (a) represents the initial condition for a rectangular constriction and (b) represents the later state. (c,d) represents the same for a sinusoidal constriction.} 
   \label{problem_description}
\end{figure}

Figure \ref{problem_description} illustrates the configurations investigated in this study. In the first configuration, the shock wave interacts with an abrupt, rectangular, localized constriction of the conduit, as shown in figure \ref{problem_description}(\textit{a}) and (\textit{b}). The second configuration considers a smooth, localized constriction, enabling a direct comparison between sharp and gradually contoured sinusoidal geometries, as in figures \ref{problem_description}(\textit{c}) and (\textit{d}). The geometries are two-dimensional, featuring a symmetric converging-diverging section with a uniform conduit height, $H$, on either side of the constriction. Both geometries permit variation in blockage ratio and length. The rectangular obstacle introduces sharp leading and trailing edges, whereas the sinusoidal contour imposes gradual curvature changes that influence boundary-layer separation. The use of these two configurations allows us to systematically examine how the constriction height, $h$, and length, $L$, govern the strength and structure of both the reflected and transmitted shock waves, as well as the evolution of the impulsively driven flow field within the constriction. Both geometries are fully defined by these two parameters, thereby limiting geometric variability while still enabling independent control of the blockage ratio, $\text{BR} = 2h/H$, and the normalised length, $\tilde{L} = L/H$. For clarity, all normalised quantities are denoted by $\tilde{\cdot}$; unless otherwise specified, lengths are normalised by the conduit height $H$ and pressures by the ambient pressure $P_0$.

In the studied scenario, a normal shock wave of moderate strength ($M_I$= 1.4 \& 1.8) propagates through initially quiescent atmospheric air (marked as state 0) and produces an abrupt jump in pressure, density, temperature, and velocity, to state~1 immediately behind the incident shock. Upon reaching the localized constriction, the shock wave interacts with the geometry and splits into a reflected shock that propagates upstream with a Mach number \(M_R\), and a transmitted shock that continues downstream with a Mach number \(M_T\). The reflected shock further compresses the flow behind the incident shock, generating state~2. Downstream of the constriction, the transmitted shock expands and, after some distance, resumes a normal, uniform propagation that is weaker than the incident shock due to energy partitioning and geometric effects, forming state~3.

The sinusoidal contour of the constriction used in this study is:
\begin{equation}
	f\left(x\right)=\frac{h}{2}\left(1+\sin\left(\frac{2\pi x}{L}-\frac{\pi}{2}\right)\right) \quad \text{for}\quad 0\leq x\leq L.
\end{equation}
This geometry has been chosen since it provides a smooth transition from the straight conduit . Hence, this geometry isolates the geometric effects on shock propagation through the constriction without introducing the complexity of a sharp discontinuity along the shock path. The smooth transition prevents spurious reflections that do not arise from the constriction itself, thereby obscuring the results. While the rectangular geometry always provides a straight-angle wall section facing the incident shock, the maximum slope of the sinusoidal geometry varies with different blockage and length ratios. It can be calculated by taking the derivative of the profile at $L/4$, resulting in a maximum slope of $\pi h/L$, and maximum angle $\theta_{\max} = \tan^{-1}\!\left(\pi h/L\right)$.

All of the cases examined in this study used a square tunnel with $H$ = 40~mm. The test matrix for the geometries examined in this study is summarized in table \ref{tab:geometry}, which lists the blockage ratio, $h/L$, and the maximum surface-slope angle at $L/4$. The minimum blockage ratio used in this study is 0.35, corresponding to $\tilde{h}=0.175$ ($h=7\text{mm}$), which is above the minimum blockage required to achieve choked flow in the constriction. This has been calculated for sinusoidal flow, assuming isentropic flow in the upstream section of the contraction at late times. Appendix \ref{app:choking} presents the calculation showing that under these assumptions, the flow becomes choked when $\tilde{h}$=0.15 ($h\sim$=6~mm), and a numerical simulation of the local Mach number along the centerline of the constriction at different times after shock wave impingement, validating that for $\tilde{h}$=0.125 ($h$=5~mm) the flow is not choked, but for $\tilde{h}$=0.15, it is.

\begin{table}
\centering
\caption{Geometric parameters of the constriction section for the three constriction lengths considered. 
}
\label{tab:geometry}
\vspace{6pt}
\setlength{\tabcolsep}{6pt}
\renewcommand{\arraystretch}{1.15}

\begin{tabular}{c c cc cc cc}
\toprule
\underline{$\tilde{h}$} & \underline{$\mathrm{BR}$}
& \multicolumn{2}{c}{\underline{$\tilde{L}=0.25$}}
& \multicolumn{2}{c}{\underline{$\tilde{L}=1$}}
& \multicolumn{2}{c}{\underline{$\tilde{L}=2$}} \\
& 
& $h/L$ & $\theta_{\max}({}^\circ)$
& $h/L$ & $\theta_{\max}({}^\circ)$
& $h/L$ & $\theta_{\max}({}^\circ)$ \\
\midrule
0.175 & 0.35
& 0.70   & 65.5
& 0.175  & 28.8
& 0.4375 & 15.4 \\
0.25  & 0.50
& 1.00   & 72.0
& 0.25   & 38.0
& 0.125  & 21.4 \\
0.30  & 0.60
& 1.20   & 75.0
& 0.30   & 43.0
& 0.15   & 25.0 \\
0.375 & 0.75
& 1.50   & 78.0
& 0.375  & 49.7
& 0.20   & 30.5 \\
\bottomrule
\end{tabular}

\vspace{4pt}
\parbox{\linewidth}{\footnotesize
*Rectangular models with similar geometric features were also considered; however, for all such cases the interaction angle was fixed at $90^\circ$ on both sides.\\
** All quantities are normalised by the conduit height $H$=40(mm) .
}
\end{table}

When a shock wave encounters a localized constriction, it induces a complex sequence of wave interactions and unsteady flow phenomena, including shock reflection at the entrance to the constriction, shock focusing or standing shock formation within the throat, and trailing expansion waves. The precise nature and intensity of these features depend not only on the incoming shock Mach number but also on the shape and extent of the constriction. While short constrictions act as abrupt obstructions, promoting strong initial reflections and jet-like flow downstream, longer transitions allow the flow to adapt more gradually, producing weaker reflections but generating complex shock structures inside the narrowed region. The blockage ratio strongly influences the strength of the upstream reflection and of the transmitted shock.

To isolate the effects of length and blockage ratio, we perform a series of controlled numerical simulations using a high-order compressible Navier-Stokes solver with large-eddy simulation (LES) modeling \citet{chandravamsi2023application, chandravamsi2024high}. The simulated flow field is validated against experimental observations using time-resolved schlieren imaging from the shock tube facility. Our goal is to develop a mechanistic understanding of how geometry governs the partitioning of wave energy between reflected and transmitted fronts and to lay the groundwork for a predictive framework linking shock properties and conduit design to wave attenuation and downstream loading.

The analysis presented in the next sections is structured to first resolve the short-time start-up processes that occur immediately after the impingement of the incident shock on the constriction, and then examine how these transient dynamics relate to the stabilized reflected and transmitted shock waves propagation in later times. We begin by investigating the transient response within and near the constriction using time-resolved flow visualizations and centerline pressure distributions, allowing us to identify the reflection mechanisms, wave interactions, and internal flow adjustments that govern the early redistribution of compression across the constriction. Once these processes subside and the flow approaches a quasi-steady configuration, the reflected and transmitted shocks propagate away from the constriction with well-defined strengths and structures that can be systematically related to blockage ratio and constriction length. The results are presented by comparing the rectangular and sinusoidal cases, which represent two extreme cases of shock-structure interaction.

\section{Methods}

\subsection{Large eddy simulations}
\label{Large eddy simulations}
This study is primarily based on numerical data obtained from our in-house large-eddy simulation solver for the three-dimensional compressible Navier-Stokes equations that uses a conservative low-dispersion finite-difference framework. The solver has previously demonstrated accuracy in resolving shock-wave propagation in conduits with abrupt area variation \citep{gichon2024dynamics}. Inviscid fluxes are computed using the sixth-order Optimized Upwind Reconstruction Scheme (OURS6) of \citet{chandravamsi2023application}, combined with the monotonicity-preserving limiter of \citet{suresh1997accurate} for shock capturing. Conservative fluxes are obtained using the HLLC approximate Riemann solver \citep{harten1983upstream,Toro1997}. Viscous fluxes are discretized using the sixth-order midpoint-based explicit optimized scheme (ME6-Opti) proposed by \citet{chandravamsi2024high}. The scheme provides low dispersion and controlled damping at high wavenumbers, which suppresses odd-even decoupling while preserving the dynamically relevant scales. Time integration is performed with the third-order total variation diminishing Runge-Kutta scheme of \citet{gottlieb1998total}. Subgrid-scale effects are represented implicitly through the intrinsic numerical dissipation at high wavenumbers, consistent with the implicit LES paradigm \citep{ahn2021numerical,chandravamsi2023application}. The solver is formulated in curvilinear coordinates with multi-block grids to accommodate complex geometries and has been extensively validated for a broad class of compressible flows \citep{chandravamsi2023application,kakumani2023gpu}. Temperature-dependent viscosity is modeled using Sutherland’s law \citep{sutherland1893lii}.

The geometry and initial conditions of the problem were chosen based on the experimental set-up. The computational domain includes only the driven and test sections, excluding the driver sections. All the simulations were performed under a grid resolution of 30 million cells. Grid clustering was employed near the wall and the entrance region of the test section with $x_{min} = y_{min} = 2 \times 10^ {- 6} h$ to capture the fine scales. Although the grid resolution employed is not fully sufficient to capture all the turbulent scales, the purpose of the present numerical simulations was restricted to accurately capturing the shock structure evolution and its downstream propagation. All the walls in the computational domain are imposed with the adiabatic wall boundary condition. At the entrance and exit of the geometry, characteristic non-reflecting boundary conditions \citep{poinsot1992boundary} are employed to ensure no reflections enter the duct. 

\subsection{Experimental system}
\begin{figure}
   \centerline{\includegraphics[scale=0.8]{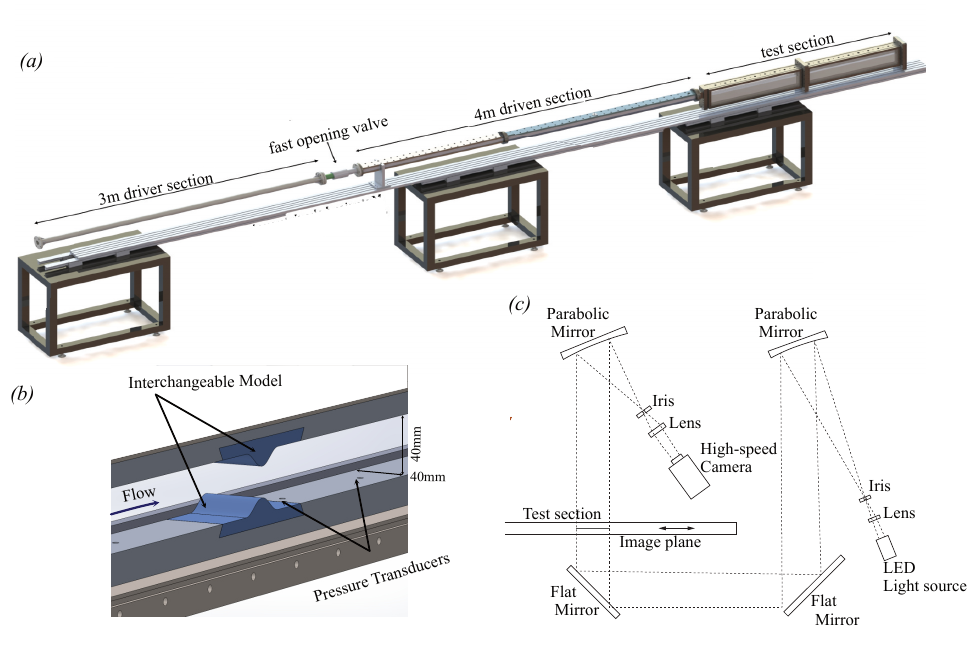}}
   \caption{Schematics of the experimental system (a) The shock tube apparatus (b) the interchangeable constriction model installed in the test section, and (c) the optical setup used to perform high-speed schlieren imaging.}
   \label{system}
\end{figure}
Experiments are conducted in the Transient Fluid Mechanics Laboratory (TFML) shock tube at the Technion - Israel Institute of Technology. Figure \ref{system}(\textit{a}) depicts the shock tube apparatus designed with a $3000~\mathrm{mm}$ long, $52.5~\mathrm{mm}$ inner diameter driver section and a $4080$ mm long, $40 \times 40$ mm$^2$ cross-section driven section. These dimensions allow us to perform experiments with constant inlet properties to the test section for 3-6 ms, depending on the Mach number.

The shock tube is connected to a square test section that is $1000~\mathrm{mm}$ long and $40~\mathrm{mm}$ wide, fitted with an interchangeable, symmetric blockage insert. A partial model of the test section is shown in figure \ref{system}(\textit{b}). The models were made from rigid plastic and held in place via a dovetail groove machined into the aluminum wall of the test section, ensuring minimal reflections along the connecting line. This modular section enables systematic measurements of shock propagation through a localized constriction with controllable geometry, blockage ratio, and streamwise length. The test section is fitted with two 40 mm thick clear acrylic windows that enable unobstructed optical access to record the flow field and shock waves evolution. Pressure is recorded in the shock tube from 8 flush-mounted transducers (Endevco 8530B-500 and 136 signal conditioners) using four oscilloscopes (LeCroy 3034Z) at 250 kHz. Two transducers are located 1000 mm apart upstream of the test section entrance to calculate the inlet incident shock wave Mach number. The imaging and the data acquisition systems are synchronised using an external timing box (Quantum composers 9400) triggered by the shock wave arrival at the first pressure transducer mounted before the test section. High-speed schlieren imaging is used to validate the numerical data. 

A schematic of the schlieren setup is presented in figure \ref{system}(\textit{c}). An LED light (Thorlabs CWHL5-C1) is directed through a pinhole to reduce non-uniformity in the beam and collimated using a parabolic mirror (Edmund Optics \# 32-276-533) with an unobstructed diameter of 286 mm. The collimated beam is aligned using two flat optical mirrors so that the light propagates perpendicularly through the test section. The collimated beam is focused using a similar second parabolic mirror through an iris at its focal point. The iris opening is set to create a schlieren image sensitive enough to capture the density variations in the flow that form a downstream of the constriction. An imaging lens is used to form the image of the test section at the plane of the camera sensor (Phantom V2640), enabling recording of the flow field at up to 80 kHz with a resolution of 1024$\times$256 pixels. The air in the driven and test sections was kept at atmospheric conditions before the experiments, as the end of the shock tube remained open. We have allowed for a long time (about 10 minutes) between experiments for the condition in the driven and test sections to re-equalize with the lab air. Consecutive experiments are run with a sufficient delay to ensure that the incident shock wave propagates into quiescent air with uniform atmospheric properties. The experiments are performed without an end wall to prevent the return of a reflected shock from the end of the test section, thereby extending the duration of the experiment.

\section{Results}
\subsection{The main features of shock wave interaction with local constriction}

\begin{figure}
   \centerline{\includegraphics[scale=0.8]{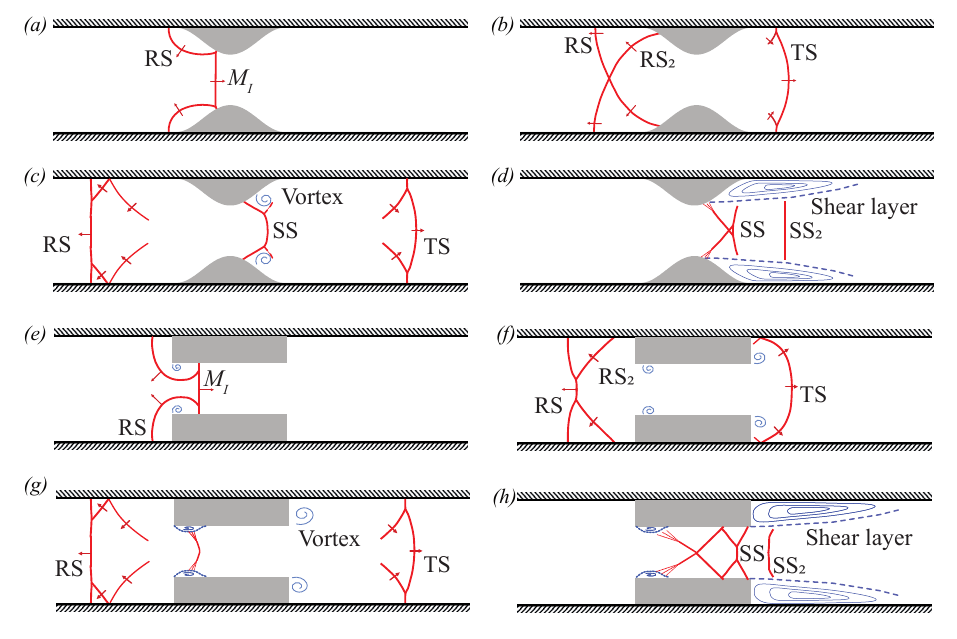}}
   \caption{A schematic description of shock wave interaction with a local constriction in a straight conduit. (\textit{a})-(\textit{d}) depicts the interaction with a smooth sinusoidal local constriction. (\textit{e})-(\textit{h}) depicts the interaction with a sharp rectangular local constriction. RS- reflected shock wave, TS - Transmitted shock wave, SS - standing shock wave } 
   \label{ref_schem}
\end{figure}

This study compares various local constriction geometries and their effects on flow and shock-wave evolution. The following description outlines the main features observed in our analysis of shock propagation, which are discussed in further detail in the next sections. A schematic description of the transient evolution of the interaction between a shock wave and a local constriction in a conduit is presented in figure \ref{ref_schem} for both smooth constriction (\textit{a-d}) and for sharp rectangular constriction (\textit{e-h}). First, as the incident shock moves into the quiescent air and reaches the constriction at a uniform velocity $M_I$, it impinges on the constriction. Immediately, the shock forms reflections at the walls, depending on the geometry. In the smooth case (figure \ref{ref_schem}(\textit{a})), the reflection always begins as a weak Mach Reflection (MR), and as the surface elevation increases, the reflection becomes stronger until reaching the maximum angle at $L/4$. The maximum slope angle depends on the specific $L$ and $h$ of the model, as shown in the table \ref{tab:geometry}. For geometries in which the angle becomes sufficiently large, the reflection will transition to a Regular Reflection (RR). The transition occurs at an angle that depends on the Mach numbers, which can be calculated for the quasi-steady case \cite{ben2007shock}, but, for the transient case, they differ significantly, with reports placing the transition angle for moderate Mach numbers of 1.2-1.8 at 52$^\circ$ to 60$^\circ$ \citep{ben2007shock,geva2013non}. In this study we performed analysis for $M_I =$ 1.4 and 1.8. To allow direct comparison between results only $M_I=1.4$ is presented. The results for $M_I=1.8$ exhibit similar trends to those of $M_I=1.4$. Reflected and transmitted shock waves speeds for $M_I=1.8$ are given in the discussion section. In the rectangular geometry case (Figure \ref{ref_schem}(\textit{e})) the shock impinges head-on onto the flat 90$^\circ$ face of the model, the reflected shock forms immediately, and a Mach reflection is formed between the incident shock that propagates in the narrow region. In this case, the induced flow behind the shock wave interacts with the sharp upstream-facing corner, leading to immediate flow separation in the form of a growing vortex. We have found that in the sinusoidal, smooth geometry, the separation of flow begins in later stages as further discussed in \S \ref{sec:transient_startup}. The shock waves that are reflected upstream expand, forming a Reflected Shock (RS) propagating upstream against the flow induced by the incident shock wave. 

As the shock pass though the throat of the constriction, it begins to expand, forming a curved expanding Transmitted Shock (TS). In the case of the sinusoidal geometry (figure \ref{ref_schem}(\textit{b}), as long as the shock wave is moving along the surface of the expanding region, it does not form a reflection. However, when the angle of the surface begins to decrease (i.e., downstream of $3L/4$), a Mach reflection forms, which persists as the transmitted shock propagates downstream. In the rectangular constriction case, the transmitted shock wave expands along the vertical walls of the downstream expansion and impinges head-on onto the walls of the conduit as depicted in figure \ref{ref_schem}(\textit{f}). Since the angle between the transmitted expanding shock wave and the wall is small, a regular reflection will form initially. Further downstream, the transmitted shock curvature will decrease, increasing the angle between the shock wave and the wall until it will also transition into a Mach reflection. A full description of this process is presented in \citet{gichon2024dynamics}. 

As the two primary shock waves i.e., the transmitted and reflected shocks, both continue to propagate away from the constrictions, the Mach reflections will continue to evolve in such a way that the triple point will reverberate between the two walls of the conduit forming multiple reflections behind them. In essence, the reflection pattern of the transmitted and reflected shock will become quite similar from this point on, as depicted in figures \ref{ref_schem}(\textit{c}) and (\textit{g}). Naturally, as the shock waves propagate away from the constriction, the repeated reflections behind them will subside, and the shock fronts will resume their normal, uniform state. The transmitted and reflected shocks will depend on the geometry and will be discussed in detail throughout the following sections. 
The geometry of the constriction will strongly affect the transient starting processes that occur within it as the primary transmitted and reflected shocks move away. At first, the reflected shocks from the transmitted and reflected shocks that form close to the constriction propagate back into it and interact with the flow, but these weaker shocks (not shown at all in the schematics of figure \ref{ref_schem}) subside rapidly. The flow in the constriction continues to evolve as the impulsively driven vortices expand, break up, and regions of flow separation form. During this time, the flow passing through the constriction accelerates, and oblique Standing Shock waves (SS) begin to form. In the sinusoidal geometry, flow separation occurs near the model's peak height, whereas in the rectangular geometry, it occurs shortly after the separation point at the upstream-facing corner. Later, as the flow in the constriction reaches a more developed state, additional shock waves form behind them, forming a normal shock train. The exact nature of this standing shock wave depends on the geometry, and they typically form downstream of the constriction when the starting vortices breakup and shear layers from that bound flow behind the constriction, as shown in figure \ref{ref_schem}(\textit{d}). In rectangular cases, they might first be preceded by a series of reflections from interactions between the first oblique shocks and the wall. In figures \ref{ref_schem} (\textit{d}) and (\textit{h}), the standing normal shocks are marked by SS$_2$

\subsection{Validation of Numerical Results}
The numerical simulations were validated against time-resolved schlieren imaging and wall-pressure measurements obtained using the shock tube facility. The validation focus on two representative geometries, both with BR=0.5 ($h$=10 mm) and $\tilde{L}=~1$ ($L$=40 mm), representing two distinctive different cases: (i) an abrupt rectangular constriction with sharp corners and immediate separation at the upstream-facing edge and an abrupt expansion at the downstream side of the constriction, and (ii) a smooth sinusoidal constriction for which the post-shock flow remains attached initially and separation develops gradually as the impulsively driven flow accelerates through the contraction. These cases, therefore, test the ability of the solver to capture both the shock morphology and the start-up flow evolution that govern the reflected and transmitted shock waves and the consequent flow within the narrow region.

\begin{figure}
   \centerline{\includegraphics[scale=0.9]{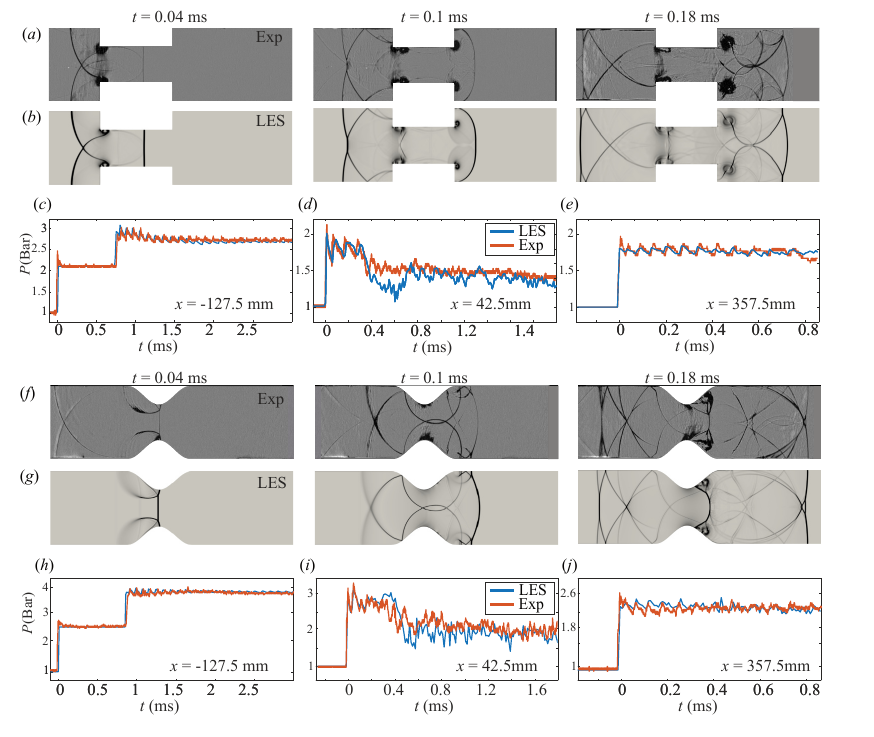}}
   \caption{Comparison of the LES results against experimental schlieren imaging and wall pressure measurements for two constriction geometries with ${h}=10mm$ (BR = 0.5) and $L$ = 40 mm ($\tilde{L}=1$).Panels (\textit{a},\textit{b}) and (\textit{f},\textit{g}) compare experimental schlieren images with numerical schlieren computed from density gradients at different times. Panels (\textit{c},\textit{d},\textit{e}) and (\textit{h},\textit{i},\textit{j}) compare LES and experimental wall pressure signals at three transducer locations: $x = -127.5\,\mathrm{mm}$ (upstream of the constriction), and $x = 42.5\,\mathrm{mm}$ and $x = 357.5\,\mathrm{mm}$ (downstream of the constriction). All pressure traces are shifted to $t = 0$ at the arrival of the incident shock at the most upstream transducer.}
   \label{valid}
\end{figure}

Figure~\ref{valid} presents side-by-side comparisons between experimental schlieren images and numerical schlieren fields constructed from density gradients, together with the comparisons of local wall-pressure histories. For the rectangular constriction, schlieren comparisons are shown in figure~\ref{valid}(\textit{a}) and (\textit{b}), while the pressure histories are shown in figure~\ref{valid}(\textit{c})-(\textit{e}). For the sinusoidal constriction, the same comparisons are shown in figure~\ref{valid}(\textit{f}) and (\textit{g}) and in figures~\ref{valid}(\textit{h})-(\textit{j}), respectively.

The schlieren comparisons in figure~\ref{valid} show that the computed shock locations and reflection patterns closely match the experimental results at each recorded time. The positions of the reflected and transmitted shocks are captured accurately, and the simulations reproduce the curvature and progressive straightening of the fronts as they propagate away from the constriction. In addition, the simulations capture the structure and evolution of the reflections between the transmitted shock waves and the wall, their transition from RR to MR configurations, the trajectory of the triple points, and the slip lines. The subsequent evolution of the reflected system, including the downstream motion of the triple points and the development of multiple reflected branches, is reproduced with high fidelity. The computed schlieren fields also capture the formation of vortical structures generated as the shock passes through the constriction, together with the ensuing shock--vortex interactions and the resulting shock pattern that emerges from them. These agreements are observed for both geometries, despite the clear difference between the sharp-cornered rectangular constriction and the smoothly contoured sinusoidal constriction.

The pressure comparisons in figure \ref{valid} provide a quantitative validation of both shock strength and timing. The upstream transducer (figures~\ref{valid}(\textit{c}) and ~\ref{valid}(\textit{h}), located at $x$ = -127.5 mm where $x$ = 0 is set to be the entrance to the constriction) upstream of the constriction, captures the incident and reflected shocks. In both geometries, the simulation reproduces the arrival time of the reflected shock and its pressure jump with very good agreement. Moreover, the simulation captures the subsequent pressure dynamics behind the reflected front, including the series of secondary compressions associated with reflections that reverberate between the conduit walls. Downstream of the constriction, the transducer closest to the model (figures \ref{valid}(\textit{d}) and \ref{valid}(\textit{i}), at $x$=42.5 mm) captures the transmitted shock and the near-field unsteady processes that occur immediately after the expansion. Farther downstream (figures \ref{valid}(\textit{e}) and \ref{valid}(\textit{j}), at $x$ = 357.5 mm), the pressure signals capture the transmitted shock after it has propagated into the uniform section and after the strongest near-field interactions have subsided. At both downstream locations, the simulation also faithfully predicts the arrival time of the transmitted shock, the magnitude of the primary pressure rise, and reproduces the pressure fluctuations induced by the reflection system that propagates behind the transmitted shock. The agreement is particularly important at the far-downstream and upstream transducers, indicating that the simulation correctly captures both the attenuation of the reflection patterns and the transient flow field evolution toward a quasi-steady mean pressure at late times.

Minor discrepancies between simulation and experiment occur at the near-downstream location $x$ = 42.5 mm shown in figures~\ref{valid}(\textit{d}) and ~\ref{valid}(\textit{i}). This region is very close to the constriction and hence most sensitive to the details of the developing separated shear layers and to three-dimensional turbulence generated by shock--vortex and shock--flow interactions, vortex breakdown, and shear layer formation. The present computations are performed as an implicit LES and therefore do not fully resolve the smallest turbulent scales; accordingly, some differences in the local pressure trace are expected. Importantly, these differences are confined to the immediate vicinity of the constriction and have a small effect on the conclusions drawn from the simulations. The predicted reflected and transmitted shock strengths and their propagation speeds in the uniform upstream and downstream sections, which underpin the quantitative trends and regime distinctions reported below, are accurately captured, as demonstrated by the very good agreement at the upstream and downstream transducers.

Beyond validating shock strength and timing, figure~\ref{valid} also shows that the simulations reproduce the formation and evolution of the local reflections and shock structures forming within the throat region. In addition, the simulations capture the formation of standing shocks downstream of the constrictions within the jet as it emerges from the constriction, and the gradual rearrangement of the flow as it reaches steady conditions. Additional validation cases were examined across both shorter and longer models and multiple blockage ratios, and they all exhibit similar agreement: close correspondence in shock locations and reflection topology in schlieren, and accurate prediction of upstream and downstream pressure histories. This establishes the reliability of the simulations for the parametric analysis presented in the subsequent sections. Further validation of the numerical framework, including its ability to capture shock propagation, reflection patterns, and pressure evolution in conduits with abrupt area changes, has also been presented in the recent work of \citet{gichon2024dynamics}, where the same solver and modeling approach were shown to reproduce experimental measurements with comparable accuracy.

\subsection {The transient startup process at the narrow region}
\label{sec:transient_startup}

\begin{figure}
   \centerline{\includegraphics[scale=0.85]{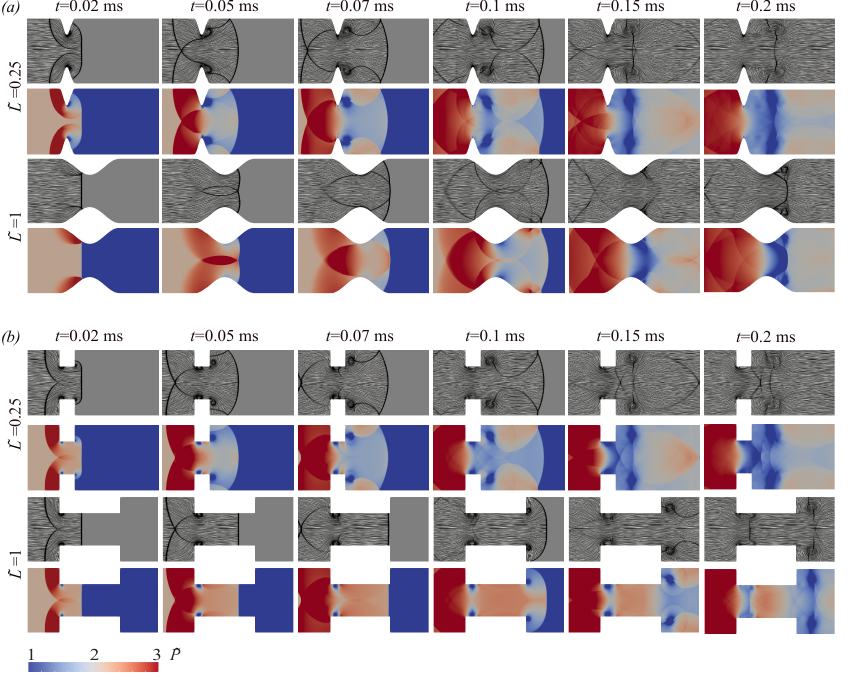}}
   \caption{The transient evolution of the shock-flow interaction within localized constrictions for cases with BR = 0.5, and different normalised lengths, $\tilde{L}=0.25,1$. For each case, line integral convolution (LIC) visualizations (top rows) illustrate the instantaneous flow structure, while the corresponding pressure fields (bottom rows) show the spatial pressure distribution.} 
   \label{Transient_compare}
\end{figure}

Figure \ref{Transient_compare} presents line integral convolution (LIC) visualizations and pressure fields obtained from the numerical simulations for a representative set of cases. The figure compares configurations with BR = 0.5  for (\textit{a}) sinusoidal and (\textit{b}) rectangular narrow-section geometries of length $\tilde{L}=0.25$ and $\tilde{L}=1$, highlighting the transient evolution of the flow field within the constriction during the short time interval following the passage of the incident shock. 

As described in figure \ref{ref_schem}, when the shock impinges on the upstream-facing wall of the rectangular constriction, it immediately forms a reflected shock wave that expands upstream while a portion of the shock propagates into the constriction. In the shorter case ($\tilde{L}=0.25$), the shock quickly exits the constriction, expands into the downstream conduit, impinges on the walls, and forms an RR. Shortly thereafter, as the shock expanded into the downstream region, the reflection patterns transitioned to an MR. A similar course of events occurs in the longer rectangular, with the only difference being that the transmitted shock propagates within the straight constriction for a longer duration. Moreover, examining the reflected shock pattern shows that its evolution during the first $\sim$0.1ms after the reflected shocks exit the image is identical in both cases. Figure \ref{Transient_compare}(\textit{a}) also shows that the flow field that forms within the constriction immediately separates at the sharp corner of the leading edge, forms a small separated region inside the constriction, and a second separation at the trailing edge of the constriction. The high-pressure gradient (discussed in detail in the next sections) that develops across the constriction induces the formation of a standing shock wave. Here, we begin to see significant differences between the cases; in the shorter case, the standing shock wave forms downstream of the constriction and is oblique, whereas in the longer case, it forms within the constriction and is a normal shock. 

Figure \ref{Transient_compare}(\textit{b}) shows that in the cases of a sinusoidal constriction, the evolution of the flow field heavily depends on the constriction length, from the very first moments of the interaction. The length of the constriction determines the rate at which the slope of the wall changes and the maximum angle of the wall, which is reached at $L/4$ (see table \ref{tab:geometry}). In the shorter sinusoidal case presented in figure \ref{Transient_compare}(\textit{b}) (BR = 0.5, $\tilde{L}=0.25$), the slope angle increases rapidly and reaches a maximum slope of $\theta_{max}=72^\circ$. As expected, the reflection pattern that forms is quite similar to the short rectangular case shown in figure \ref{Transient_compare}(\textit{a}), exhibiting an immediate formation of a strong reflection expanding upstream. However, the gradual slope angle increase and the shallower maximum angle of the second case presented in figure \ref{Transient_compare}(\textit{b}) ($\text{BR} = 0.25, \tilde{L} = 1$, $\theta_{max} = 38^\circ$) leads to the formation of a weaker Mach reflection that is quickly dispersed into a series of pressure waves.

\begin{figure}
   \centerline{\includegraphics[scale=0.80]{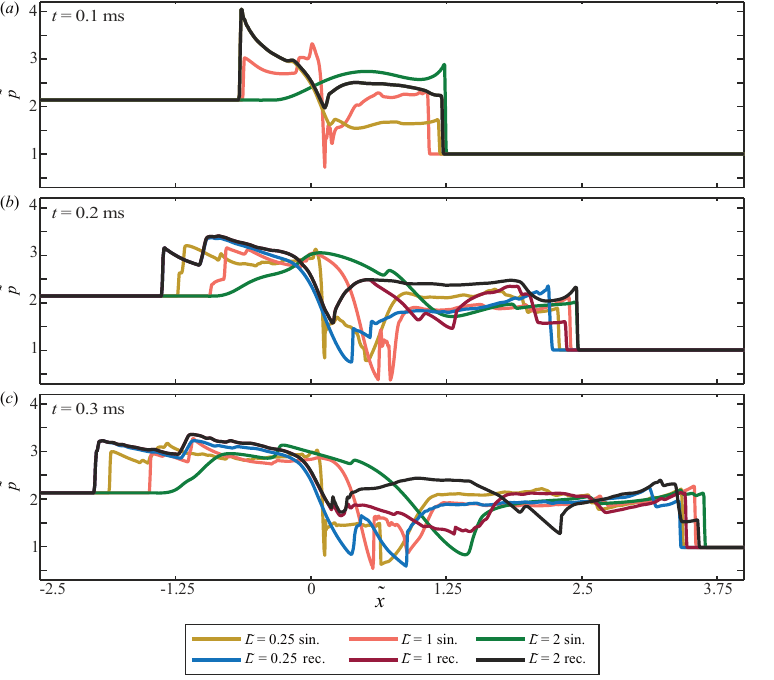}}
   \caption{Centerline pressure distributions in the vicinity of the constriction during the early-time following the impingement of the incident shock wave with $M_I = 1.4$. Profiles are shown over the domain $\tilde{x}=-2.5$ to 4 at three successive times: (\textit{a}) $t = 0.1~\mathrm{ms}$, (\textit{b}) $t=0.2~\text{ms}$, and (\textit{c}) $t = 0.3~\text{ms}$. The results compare sinusoidal and rectangular constrictions with identical blockage ratio and three lengths ($\tilde{L}=0.25,\ 1,\ 2$). 
} 
   \label{Transient_compare_pressure}
\end{figure}

Figure \ref{Transient_compare_pressure} presents the instantaneous normalised pressure distribution along the conduit centerline in the vicinity of the constriction, comparing the short-time evolution of the pressure field for rectangular and sinusoidal cases with a blockage ratio of BR = 0.5. The figure presents LES results for various model lengths following the impingement of an incident shock wave with $M_I=1.4$. The profiles capture the developing reflected and transmitted shock systems, as well as the pressure evolution within the constriction shortly after being impulsively driven by the passage of the incident shock wave.

At the earliest time shown ($t=0.1~\mathrm{ms}$), the pressure distribution generated by all rectangular cases is virtually identical, indicating minimal effects of the constriction length in early times. Conversely, in the case of a sinusoidal geometry, differences in pressure distribution occur immediately, as one would expect, given the fact that the impinging shock wave interacts with geometries of different slopes. Figure \ref{Transient_compare_pressure} shows that for the shortest sinusoidal model ($\tilde{L}=0.25$), a strong reflection occurs that is identical to the rectangular cases, but the transmitted shock wave is significantly reduced in strength. The longer case of $\tilde{L}=1$ creates a weaker reflected shock wave and a stronger transmitted shock wave, while the longest case with $\tilde{L}=2$ doesn't form a reflected shock wave at this time, but the transmitted shock wave is strong and still propagating within the constriction.

As time progresses, the reflected shock waves from the rectangular geometries remain approximately constant in strength, indicating that reflected shock strength is independent of the constriction length. However, the transmitted shock waves begin to exhibit geometric effects arising from differences in reflection patterns and flow fields within the constrictions. For the sinusoidal cases, figures \ref{Transient_compare_pressure}(\textit{b}) and (\textit{c}) show that the evolution of the reflected shock wave depends strongly on the geometry of the constriction. For the case of $\tilde{L}=2$, we see that while steepening of the pressure waves propagating upstream occurs, it has not yet formed a shock wave.

\begin{figure}
   \centerline{\includegraphics[scale=0.80
]{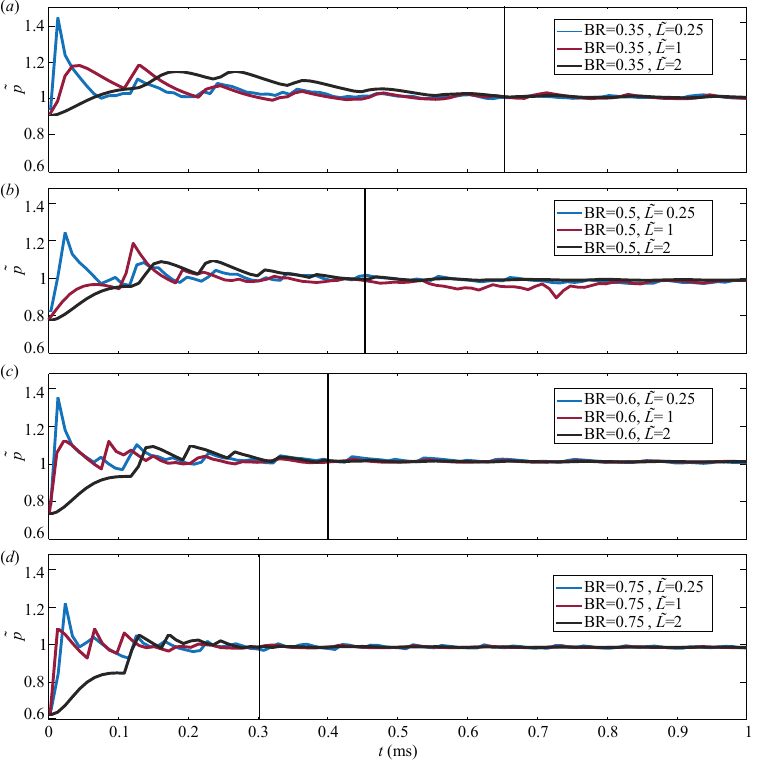}}
   \caption{Time histories of the normalised centreline pressure at the entrance to the constriction, \(x = 0\), for sinusoidal constrictions. The pressure is normalised by \(P_2\), the late-time pressure immediately upstream of the constriction (state 2). Each panel corresponds to a fixed blockage ratio BR, and compares three lengths \(\tilde{L} = 0.25,\ 1,\ 2\). The traces are shown after the passage of the incident shock (state 1), and illustrate the duration of the start-up process required for the flow to approach its late-time pressure behind the reflected shock wave $P_2$. The vertical line in each panel marks an estimated end of the transient, determined by visual inspection.}
   \label{time_to_ss}
\end{figure}

The differences in the early-time pressure response between the sinusoidal cases are directly linked to the reflection topology imposed by the local contour slope, as shown in Figure \ref{Transient_compare}. In the shortest sinusoidal constriction, the large maximum surface angle ($\theta_{\max}$) forces the incident shock to undergo a regular reflection, producing a strong, spatially coherent reflected shock similar to that observed in the rectangular geometries. As the constriction length increases, the maximum slope decreases, and the reflection persists as a Mach reflection over much of the contour. In this regime, compression is more distributed, causing the upstream response to initially appear as a train of compression waves that steepen and coalesce into a fully developed reflected shock only at later times. The multiple pressure jumps observed within and downstream of the constriction arise from repeated reverberations between these evolving reflection structures and the impulsively accelerated, separating flow within the constriction, as also evidenced by the separation and shear-layer development discussed in figure~\ref{Transient_compare}.

The early-time reflection dynamics ultimately govern the starting time of the flow in the constriction. Eventually, as the reflected and transmitted shock waves propagate away from the constriction, the strong reflections within the constriction will subside, and the flow will reach steady conditions on the upstream and downstream sides, marked by states 2 and 3 in figure~\ref{problem_description}, respectively. As this process proceeds, the reflected and transmitted shocks evolve toward well-defined steady propagation, providing the basis for examining their late-time strengths.

To highlight the temporal effects of geometry on the flow starting time, figure \ref{time_to_ss} shows the temporal evolution of the centerline pressure at the entrance to the constriction ($x = 0$) for the sinusoidal geometries. The pressure is normalised by $P_2$. Note that since $P_2$ depends on the blockage ratio, the normalised starting pressure at $t = 0$ differs across cases, whereas in reality it is the same for all ($P_1$). This figure shows the time required for the pressure at the constriction entrance to reach its late-time steady level.

Figure \ref{time_to_ss} shows that more constricted geometries lead to shorter starting times for all lengths, and that for a given blockage ratio, sharper geometries produce stronger reflections that further reduce buildup times but increase pressure fluctuations. While this might seem surprising at first, it can be intuitively explained by the fact that shock waves moving through the constriction induce rapid changes in the flow configuration, thereby promoting faster adjustments of the flow to the geometry. Vertical black lines are plotted on Figure \ref{time_to_ss} to show the time at which the pressure within the longest constrictions, i.e. $\tilde{L} = 2$, roughly reaches steady state pressure. Because the signals exhibit weak residual oscillations, an exact settling time cannot be accurately defined. We therefore estimate the end of the transient by visual inspection. The start-up process within the narrowing occurs on timescales that are approximately one to two orders of magnitude longer than the shock passage time. For $\tilde{L}=2$, the $M_I$ = 1.4 shock wave passes through the constriction in $\sim165 \mu\mathrm{s}$  and for $\tilde{L}=0.25$ the shock passes through the constriction in just $\sim 20 \mu \mathrm{s}$. The largest discrepancies occur for the least constricted cases (BR = 0.35), where the flow field remains transient with strong fluctuations for as much as $\sim 650 \mathrm{\mu s}$. As upstream processes subside, the flow field in the constriction reaches a relatively stable state, in which the pressure gradient across the constriction corresponds directly to the strengths of the reflected and transmitted shock waves, as further examined in the following section.

\subsection{The late-time evolution of the reflected and transmitted shock waves }
\subsubsection{Rectangular constrictions}

\begin{figure}
  \centering
  \includegraphics[width=0.95\linewidth]{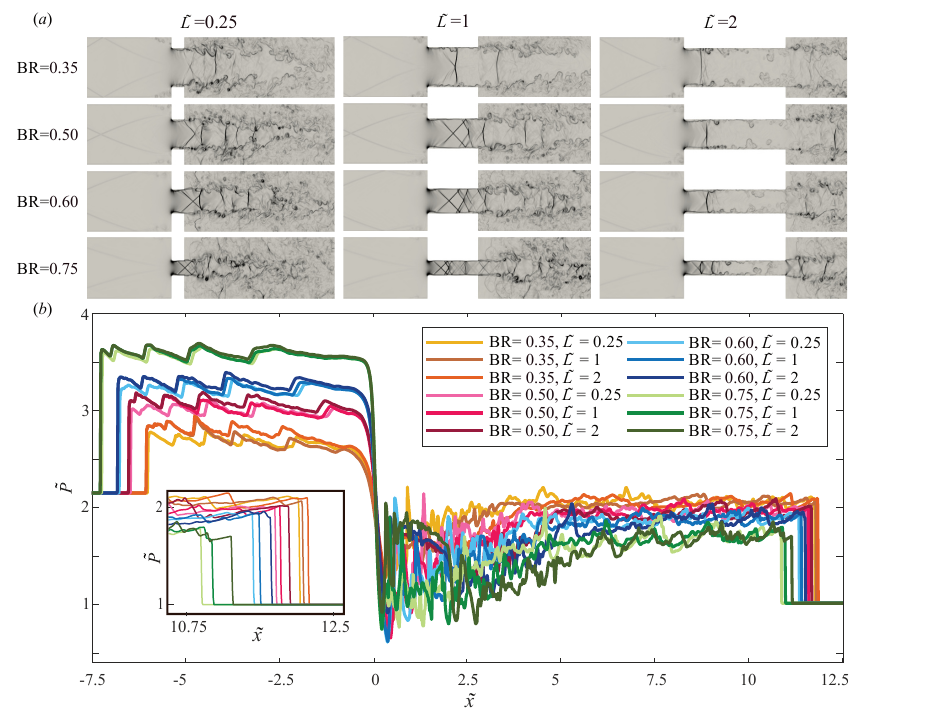}
  \caption{
  (\textit{a}) Numerical schlieren images at \(t = 1\,\mathrm{ms}\) after the initial interaction of the incident shock with rectangular constrictions. Results are shown for four blockage ratios, BR = 0.35, 0.5, 0.6, 0.75, and three constriction lengths, \(\tilde{L}=\)0.25, 1, 2.
  (\textit{b}) Pressure distribution along the centerline of the computational domain. The pressure profiles summarize the combined effects of blockage ratio and constriction length on the evolution of the transmitted and reflected shock waves for the rectangular geometries following the impingement of a $M_I$ = 1.4 shock wave. A magnified view of the transmitted shock front is shown in the inset.}
  \label{sqr_schlieren+pressure_at_t=1ms}
\end{figure}

Figure~\ref{sqr_schlieren+pressure_at_t=1ms} summarizes the late-time flow structures and centerline pressure distributions, 1 ms after the initial shock interaction with the upstream face of the rectangular constrictions. The figures show the steady-state flow conditions inside the constriction after both the transmitted and reflected shock waves have propagated away from it. Twelve representative cases are shown, corresponding to four blockage ratios (BR = 0.35, 0.5, 0.6, 0.75 ) and three constriction lengths (\(\tilde{L}\)=0.25, 1, 2). Figure~\ref{sqr_schlieren+pressure_at_t=1ms}(\textit{b}) shows the corresponding pressure distribution along the centerline of the computational domain, spanning a $20H$ section of the tunnel from \(\tilde{x}=-7.5\) to \(\tilde{x}=12.5\).

As described above, in all rectangular-geometry cases, the initial shock-constriction interaction is qualitatively identical. When the incident shock impinges head-on on the upstream-facing vertical wall of the constriction, it immediately forms a normal reflected shock sections that propagate back upstream. Simultaneously, the portion of the incident shock that passes through the opening continues into the constriction and propagates downstream. where it expands through a sharp $90^\circ$ corner. Since the inlet and outlet geometries of all the rectangular constrictions are the same, we see minor effects of the geometry length in late times, and the differences observed in figure \ref{sqr_schlieren+pressure_at_t=1ms} are driven primarily by the blockage ratio, which determines the late-time pressure gradient on the constriction. The schlieren images in figure~\ref{sqr_schlieren+pressure_at_t=1ms}(\textit{a}) show that the shock configuration that develops inside the constriction after the initial startup process has subsided is highly affected by the geometry. In all cases, the flow separates immediately at the sharp upstream-facing corners of the constriction, creating an effective contraction region. In all cases presented in figure \ref{sqr_schlieren+pressure_at_t=1ms} the flow through the constriction becomes choked, and as a consequence, shock waves form within the constriction. In some cases, we observe an oblique shock wave forming very close to the leading edge where the separated flow reattaches. This will occur if the constriction is significant, leading to a buildup of sufficient gradients across the constriction. However, if the constriction becomes longer, the resulting significant resistance to flow will slow it down at the entrance and delay the formation of a shock wave. In this case, the shock forms downstream, after the flow has adjusted to the geometry of the narrow region, resulting in a standing normal shock wave. Figure \ref{sqr_schlieren+pressure_at_t=1ms}(\textit{a}) show that when BR = 0.35, normal shocks form in all cases, but as the blockage increases to cases where BR = 0.5, in the shorter cases $\tilde{L}$=0.25 and $\tilde{L}$=1 oblique shocks are formed, but not for $\tilde{L}$=2. For cases with $\tilde{L}$= 2, we only see clear oblique shocks form in the most constricted case where BR = 0.75.   

Examining the pressure distributions presented in Figure~\ref{sqr_schlieren+pressure_at_t=1ms}(\textit{b}) shows that for a given blockage ratio, the reflected shock structure is largely insensitive to the length of the constriction. Again, highlighting that the initial shock-constriction interaction is fundamentally identical across all rectangular geometries with the same blockage ratio, the reflected shock strength depends primarily on the blockage ratio rather than on $\tilde{L}$. This behavior is expected and consistent with physical intuition, as the reflected shock is generated immediately by the head-on interaction of the incident shock with the upstream-facing vertical wall of the constriction. The agreement across different lengths further reinforces that the numerical simulations capture the underlying physics consistently and in a physically meaningful manner.

As expected, the transmitted shock shows a stronger dependence on both the blockage ratio and the constriction length, with the most open configurations generating the strongest transmitted shocks and increasing blockage leading to weaker downstream propagation. Superimposed on this trend, however, is a clear and systematic influence of constriction length. For a given blockage ratio, shorter constrictions consistently produce weaker transmitted shocks than longer ones. The magnified inlay plot in figure \ref{sqr_schlieren+pressure_at_t=1ms}(\textit{b}) shows the difference in the transmitted shock fronts more clearly. This effect is most evident in the more constricted configurations and can be understood by examining flow evolution within the constriction, as revealed by the schlieren images. Since longer constrictions allow better pressure matching between upstream and downstream, with fewer shocks, the flow exits the constriction as a stronger, more coherent jet, thereby supporting a stronger transmitted shock downstream. 

The reflected shock wave strength does not depend on the length of the narrow rectangular section, yet small differences are observed in the pressure field trailing the reflected shock as it propagates upstream. These differences become more pronounced as the blockage ratio increases. These differences can be associated with the interaction between the shock structures forming within the constriction and the reflected shock as it builds up and propagates upstream. In the most open cases, weaker reflections lead to a slower buildup of a fully developed normal reflected shock, allowing additional interaction between the internal shock system and the upstream-propagating reflected wave. This interaction influences the subsequent shock train that develops behind the reflected front. Aside from these minor variations, the reflected shock speed and the reverberations trailing it are governed almost exclusively by the blockage ratio.

\subsubsection{ Sinusoidal constrictions}
\label{sec:RedVsTrans}
\begin{figure}
  \centering
  \includegraphics[width=0.95\linewidth]{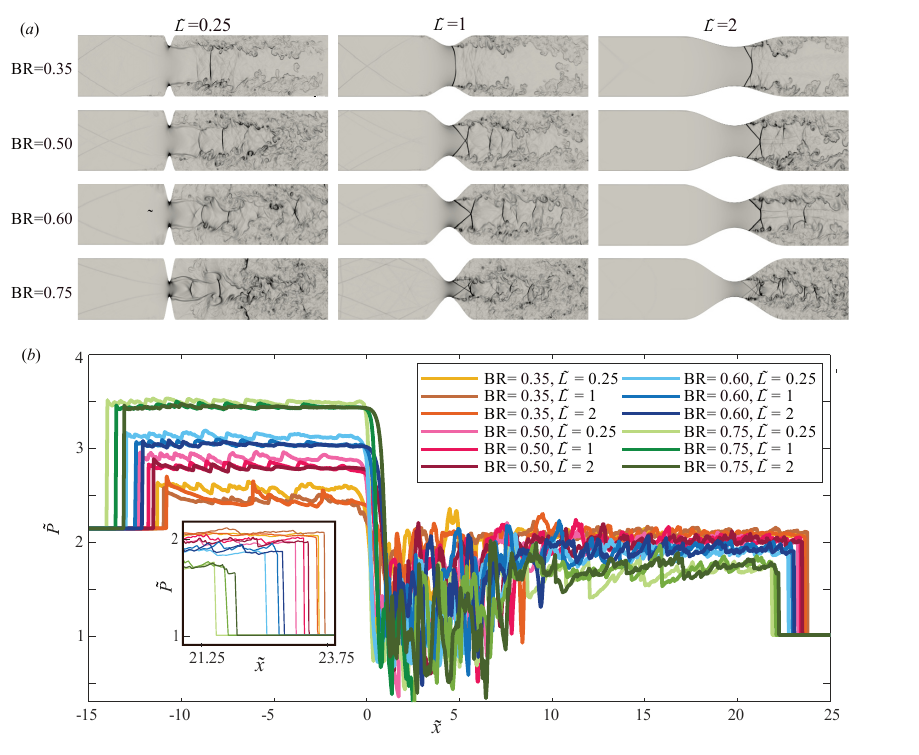}
  \caption{
  (\textit{a}) Numerical schlieren images at \(t = 2\,\mathrm{ms}\) after the interaction of the incident shock with sinusoidal constrictions. Results are shown for four blockage ratios, BR = 0.35, 0.5, 0.6, 0.75, and three constriction lengths, \(\tilde{L}\)=0.25, 1, 2.
  (\textit{b}) Pressure distribution along the centerline of the computational domain over a \(40H\) section of the tunnel, from \(\tilde{x}=-15\) to \(25\,\). The pressure profiles summarize the combined effects of blockage ratio and constriction length on the evolution of the transmitted and reflected shock waves for the sinusoidal geometries following the impingement of a $M_I$=1.4 shock wave. A magnified view of the transmitted shock front is shown in the inset.
  }
  \label{sin_schlieren+pressure_at_t=2ms}
\end{figure}
\begin{figure}
   \centerline{\includegraphics[scale=0.85]{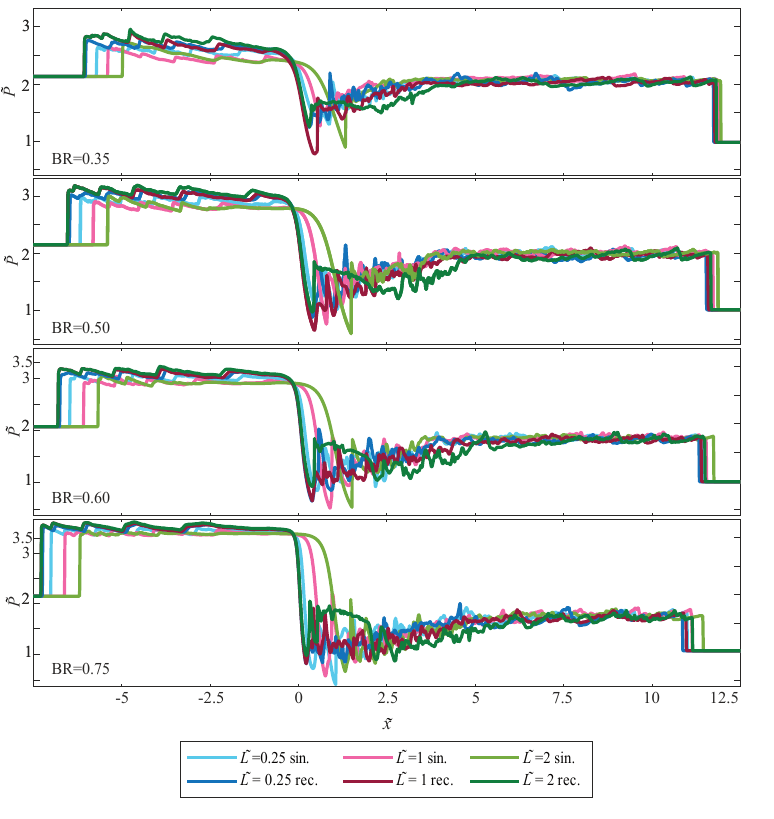}}
   \caption{A comparison of the centerline pressure distribution comparing the late time shock wave between sinusoidal and rectangular constrictions after the impingement of an $M_I=1.4$ shock wave, $t=1$ms after the shock enters the constriction.}
   \label{sqr+sin_h=10_L=10,40,80_t=1ms}
\end{figure}
Figure~\ref{sin_schlieren+pressure_at_t=2ms} compares numerical schlieren images and centerline pressure profiles for shock interaction with sinusoidal constrictions, using the same format as figure~\ref{sqr_schlieren+pressure_at_t=1ms} for the rectangular geometries. Results are shown at $t = 2~\text{ms}$, after the flow within the constriction has reached a quasi steady configuration and the dominant transients associated with the initial shock-geometry interaction have largely subsided. At these late times, the overall flow topology resembles that of the rectangular cases, but the smoother contraction and expansion in the sinusoidal geometries modify the internal shock structure. For the short, constrictions ($\tilde{L}=0.25$), no standing shocks form within the constriction. Instead, the shock pattern appears only downstream, within the core of the emerging jet. For longer constrictions ($\tilde{L}=1$), standing shocks develop within the constriction, and the configuration depends on the constriction: a nearly normal shock is observed for the most open case, whereas more constricted cases exhibit an oblique shock system. Similar trends persist for $\tilde{L}=2$, in contrast to the rectangular geometries, where the internal shock configuration shows a stronger tendency toward a nearly normal shock within the constriction. 

Examining the centerline pressure traces in figure~\ref{sin_schlieren+pressure_at_t=2ms}(\textit{b}) shows that the transmitted shock strength is primarily governed by the blockage ratio, consistent with the rectangular cases. For the sinusoidal constrictions, however, the dependency on geometric details, both in the shock arrival time and in the pressure level behind the transmitted front, is noticeably weaker. This is consistent with the schlieren fields at late times, which indicate that the internal flow configuration is broadly similar across cases at a given blockage ratio. In addition, the smoother contraction and expansion allow the transmitted wave to exit the constriction through a gradual difference in the downstream evolution. However, comparing the pressure plots in figure \ref{sin_schlieren+pressure_at_t=2ms}(\textit{b}) to the rectangular cases reveals that, in the sinusoidal cases, the reflected shock wave is highly affected by the geometry. This effect is significant in all of the cases we have examined. In essence, these differences stem from the fact that the reflected shock wave does not form immediately upon its impingement on the constriction, as would have been the case in the rectangular geometry. This trend is further illustrated in figure~\ref{sqr+sin_h=10_L=10,40,80_t=1ms}, which directly compares the centerline pressure profiles for all six rectangular and sinusoidal cases with BR = 0.5 and $\tilde{L}$=0.25, 1, and 2. The comparison shows that rectangular geometries produce the strongest reflected shocks, thereby providing an upper bound on the strength of reflected shocks for a given blockage ratio. In all sinusoidal cases, the reflected shock is weaker, consistent with the smoother contraction that reduces the impulsive compression at the entrance. Within the sinusoidal cases, the reflected shock strength decreases as $\tilde{L}$ increases, indicating that a longer, more gradual constriction further weakens the upstream compressive response. Hence, it is clear that there is a strong coupling between the short-time evolution of the shock propagation through the constriction and the buildup of transmitted and reflected waves.

Taken together, the results presented in this section reveal that the interaction between a shock wave and a localized constriction is governed by a highly complex and geometry-dependent transient start-up process, involving separation, vortex formation, and multiple shock–structure interactions that evolve over time scales significantly longer than the shock passage itself. These early-time dynamics differ markedly between abrupt and smoothly contoured geometries and strongly influence the pathway by which the flow adjusts to the constriction. However, despite these pronounced differences in transient behaviour, the results consistently show that the final, late-time reflected and transmitted shock strengths collapse onto well-defined trends that depend primarily on the blockage ratio, with secondary sensitivity to geometric details.

This apparent contradiction highlights an important point: although the short-time evolution depends on the detailed physics of shock-geometry interactions, it does not provide a universal description of the system. A complete, predictive understanding of the transient dynamics would require resolving the full flow field using high-fidelity numerical simulations as done here. Nevertheless, the systematic behaviour observed at late times suggests that the essential outcome of this complex transient process can be represented through reduced-order descriptions.

\section{Discussion}
\begin{figure}
    \centering
    \includegraphics[width=0.9\textwidth]{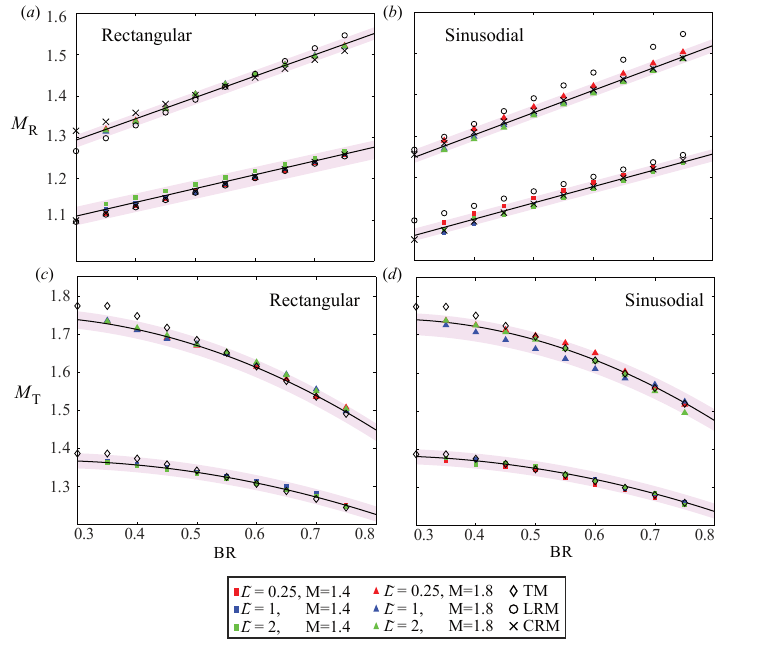}
    \caption{Reflected and transmitted shock Mach numbers as a function of the blockage ratio BR for localised constrictions of varying normalised length $\tilde{L}$. Panels~(\textit{a} \& \textit{c}) show rectangular geometries and panels~(\textit{b} \& \textit{d}) sinusoidal geometries; Reflected shocks are shown in (\textit{a} \& \textit{b}) and Transmitted shocks in (\textit{c} \& \textit{d}). Results are presented for incident Mach numbers $M_I = 1.4$ and $1.8$ and for $\tilde{L} = 0.25$, $1$ and $2$. Symbols denote numerical data and solid lines are least-squares fits to a linear function in (\textit{a} \& \textit{b}) and a parabolic function in (\textit{c} \& \textit{d}), with shaded bands indicating the $95\%$ fit uncertainty. Reduced order models predictions are also shown: $\diamond$,~TM (transmitted shock wave model); $\circ$,~LRM (linear reflected shock wave model); $\times$,~CRM (choked-flow reflected shock wave model).
}
    \label{M_R_M_T}
\end{figure}

Building on the physical understanding of flow evolution developed above, we now turn to develop a modeling framework to predict the late-time strengths of the reflected and transmitted shock waves. This approach does not attempt to resolve the full transient dynamics but instead leverages their cumulative effect to construct a simplified, predictive description of the quasi-steady shock system. Figure \ref{M_R_M_T} presents the late-time reflected and transmitted shock velocities for $M_I=1.4$ and for $M_I=1.8$ obtained from LES. Empirical least-squares fits to the LES results are added as solid lines on figure \ref{M_R_M_T} using a linear function for the reflected shock waves and a parabolic function for the transmitted shock waves. The shaded areas plotted around the fitted lines represent a 95\% confidence region. The functions and their fitting constants are provided in appendix \ref{appendix_C}. The empirical fits enable us to quantify the dependence of the reflected and transmitted shock strengths on constriction blockage ratio and incident Mach number, but they do not, by themselves, provide a mechanistic explanation for the observed behavior. At present, to the best of our knowledge, no direct analytical or semi-analytical method exists to predict the strengths of reflected and transmitted shock waves caused by a localized constriction. In the past, analytical prediction of the propagation of shock waves through regions of area change, expansion, or contraction was carried out using Geometric Gas Dynamics (GSD) \citep[e.g.][]{whitham1957new,whitham1959new,skews1967shape,ridoux2019beyond,ndebele2017propagation,ndebele2019interaction,gichon2024dynamics}. However, GSD is not well suited for describing a shock propagating through a local constriction, as it assumes free propagation that is independent of post-shock and reflected flow effects. In a constricted geometry, strong wall-induced disturbances and reflected waves interact with the shock, modifying both its strength and structure. In addition, choking within the constriction imposes a mass flow limitation that significantly weakens the transmitted shock, an effect that is not captured by GSD. A comparison of the GSD-predicted shock velocity and LES is presented in Appendix \ref{app:GSDLES}, which shows that it is inadequate to predict the shock evolution through the constriction, even for small blockage ratios. However, the regularity of the observed trends in figure \ref{M_R_M_T}, together with the dominant role of blockage ratio, provides a firm basis for developing a reduced-order framework that relates the long-time shock strengths.

\subsection{Predictive models for the Reflected shock wave}
\label{sec:model_MR}

\subsubsection{Reflected shock Linear model }

As shown in figure~\ref{M_R_M_T}(\textit{a}) and (\textit{b}), the reflected shock wave Mach number exhibits an approximately linear dependence on blockage ratio, across the full range of geometries and for both incident Mach numbers considered in this study. This near-linear collapse indicates that, to a leading order, the reflected shock strength is governed by the fraction of the conduit cross-section that experiences wall-like compression during the interaction. When the incident shock reaches the constriction, the cross-section is effectively partitioned into blocked and open regions. Over the blocked fraction, the interaction locally resembles reflection from a rigid wall, producing a compression comparable to a wall-reflection state \(M_{RW}\). Over the open fraction, the flow continues into the constricted region and does not directly contribute to upstream reflection at that instant. The reflected shock wave that ultimately propagates upstream, therefore, represents the integrated compression across these two regions.

Two limiting cases are well defined; In the limit of full blockage, corresponding to the constriction reaching the channel centerline, the reflection reduces to classical rigid-wall reflection, for which the reflected shock strength \(M_{RW}\) is uniquely determined by the post-reflected state from:
\begin{equation}
\label{eq:MRW}
\frac{M_{RW}}{M_{RW} - 1} = \frac{M_I}{M_I^2 - 1} \sqrt{1+\frac{2(\gamma - 1)}{(\gamma + 1)^2} \,(M_I^2 - 1)\left(\gamma + \frac{1}{M_I^2}\right)
}
\end{equation}
where $\gamma$ is the heat capacity ratio, which for all calculations in this study is considered to be constant, $\gamma=1.4$.

In the opposite limit of vanishing blockage, the reflection approaches a weak compression wave with \(M_R \rightarrow 1\) (sonic wave). In practice, however, even modest blockage ratios induce resistance to the flow within the constriction and generate finite pressure buildup, so the reflected Mach number remains measurably above unity throughout the investigated range. 

The observed linear dependence suggests that the reflected compression scales proportionally with the blocked fraction of the cross-section. This motivates a mixing-type Linear Reflected shock wave Model, LRM, in which the reflected Mach number varies linearly between the two limiting states,
\begin{equation}
M_R = a + \mathrm{BR}\,\left(M_{RW}-1\right)
\label{eq:reflected_shock}
\end{equation}
where the offset \(a\) accounts for the finite compression associated with choked flow and tends to be higher than unity even for low blockage ratios. Using least-squares fitting, we find that, for the cases studied here, $a = 0.99$ and $a = 1.08$ for $ M_I = 1.4$ and $ M_I = 1.8$, respectively. This formulation predicts that the reflected shock strength is primarily controlled by the blockage ratio and only weakly by the constriction length, consistent with empirical trends. The LRM predictions are plotted in figure~\ref{M_R_M_T}(\textit{a}) and (\textit{b}), and quantitative comparisons are provided in table~\ref{tab:MR_model_comparison_improved}. The predictions of LRM are compared with the LES results. The disparity between LRM and LES is calculated as $\Delta \text{LRM}\% =( \text{LRM}-\text{LES})/\text{LES}\times100$. The LES values are taken from the least-squares fit to the LES results (see appendix \ref{appendix_C}). The agreement is strongest for rectangular geometries, where the incident shock impinges directly on a vertical blocked face, and the local wall-reflection analogy is most appropriate. For rectangular cases, the mean relative RMS difference to LES is $0.83\%$ for $M_I=1.4$ and increases to $1.22\%$ for $M_I = 1.8$. 
For sinusoidal geometries, the LRM over-predicts the reflected strength, which can be explained by the fact that the compression develops progressively along the curved surface rather than instantaneously over a sharply blocked fraction. The corresponding RMS differences to LES are $2.4\%$ for $M_I = 1.4$ and $2.84\%$ for $M_I = 1.8$. The increase in the difference at the higher incident Mach number is consistent with the stronger nonlinear effects and the greater sensitivity of the reflection process to distributed geometric interactions at higher shock strength.

\subsubsection{Choked-flow model}
A more fundamental description follows from the fact that, at late times, the pressure behind the reflected shock wave is sufficiently high to produce choked flow in the constriction. Therefore, we can devise a Choked flow condition based reflected shock model, CRM, for the pressure buildup from the known jump conditions across the shock waves from which we can find $M_R$ as a function of $M_I$ and BR.
As the shock wave passes through the constriction, it induces flow conditions at the entrance with velocity $u_1$ in the lab frame of reference and a speed of sound, $a_1$, both of which are known functions of $M_I$:
\begin{equation}
  \frac{T_1}{T_0} =
  \frac{[2\gamma M_I^2-(\gamma-1)][(\gamma-1)M_I^2+2]}{(\gamma+1)^2 M_I^2},
  \qquad
  u_1 = M_I a_0 \left[1 - \frac{(\gamma-1)M_I^2+2}{(\gamma+1)M_I^2}\right],
  \label{eq:state1}
\end{equation}
Once a reflected shock is formed and propagates upstream at $M_R$, the velocity of the flow behind it, $u_2$, can be calculated from $M_R$, with respect to the lab-frame, as
\begin{equation}
  u_2 = u_1 - M_R a_1\!\left[1 - \frac{(\gamma-1)M_R^2+2}{(\gamma+1)M_R^2}
  \right].
  \label{eq:state2}
\end{equation}
By using the definition of the speed of sound for an ideal gas, we get that  $a_1=a_0\sqrt{T_1/T_0}$, from which we can find the speed of sound behind the incident and reflected shock waves as 
\begin{equation}
\label{a1}
  a_1  = a_0\,\frac{\sqrt{[2\gamma M_I^2-(\gamma-1)][(\gamma-1)M_I^2+2]}}{(\gamma+1)M_I},
\end{equation}
and 
\begin{equation}
  a_2 = a_1\,\frac{\sqrt{[2\gamma M_R^2-(\gamma-1)][(\gamma-1)M_R^2+2]}}{(\gamma+1)M_R},
\end{equation}
respectively.\\

The gas downstream of the reflected shock approaches the throat with a Mach number \(M_2 = u_2 / a_2\). As it enters the constriction, the flow is compressed from the upstream conduit area \(A_c\) toward a minimum cross-sectional area at which choking occurs, denoted as \(A^*\). Due to flow separation within the constriction, however, the effective minimum flow area is smaller than the geometric opening \(H - 2h\). This reduced cross-sectional area, referred to as the vena contracta, defines the effective choked flow area as \(A^*_{\mathrm{eff}} = C_c (H - 2h),\) where \(C_c\) is the contraction coefficient. Accordingly, an effective blockage ratio can be defined as \(\mathrm{BR}_{\mathrm{eff}} = 1 - C_c (1 - \mathrm{BR}),\) which accounts for the reduction in available flow area. The resulting effective compression from the upstream conduit to the most constricted section is therefore
\begin{equation}
\frac{A_c}{A^*_{\mathrm{eff}}} = \frac{1} { 1 - \mathrm{BR}_{\mathrm{eff}}}.
\end{equation}

The choking constraint is then simply that the flow accelerates
isentropically from $M_2$ to $M=1$ through the constriction, satisfying the
isentropic area-Mach relation 
\begin{equation}
  \frac{1}{M_2(M_R,\,M_I)}
  \left[\frac{2}{\gamma+1}\!\left(1+\frac{\gamma-1}{2}M_2^2\right)
  \right]^{\!\frac{\gamma+1}{2(\gamma-1)}}
  =
 \frac{1}{1-\mathrm{BR}_{\mathrm{eff}}}.
  \label{eq:MR_closure}
\end{equation}
Because the left-hand side, known as $f(M_2)$ is a monotonic function of $M_R$ for fixed $M_I$, the solution is unique and found efficiently by bisection. 

Since the CRM uses the isentropic area-Mach relation, \eqref{eq:MR_closure}, it implicitly assumes that entropy generation between the post-reflected state and the vena contracta is negligible. In practice, however, the flow within the constriction is far from isentropic as documented in \S\ref{sec:transient_startup}-\S\ref{sec:RedVsTrans}. The flow through the constriction involves separation at the inlet, vortex formation, shock-vortex interactions, and turbulent shear-layer development. These irreversibilities are collectively absorbed into the contraction coefficient $C_c$, which reduces the effective throat area below its geometric value and thereby accounts for the total-pressure loss along the flow path, a treatment standard in quasi-one-dimensional nozzle analysis and well-established for steady separated flows. For a rectangular constriction, classical analyses give $C_c \approx 0.61$ for an infinitely thin orifice plate \citep{lamb1924hydrodynamics} and $C_c \approx 0.75$-$0.82$ for a finite-length orifice with partial reattachment \citep{idelchik1986handbook}. It has been shown that compressibility effects can also increase the value of $C_c$ \citep{jobson1955flow}. For sinusoidal constrictions, the smooth contour suppresses immediate separation and $C_c$ approaches unity, with the small remaining deficit attributable to boundary-layer displacement at the throat. Since $C_c$ is fundamentally a geometrical coefficient, we calibrate a single value jointly over both incident Mach numbers for each geometry type, minimizing the root-mean-square percentage error across all LES data points, and find that a constant value suffices, yielding

\begin{equation}
    C_c^{\mathrm{rect}} = 0.788, \qquad C_c^{\mathrm{sin}} = 0.940.
    \label{eq:Cc}
\end{equation}

\begin{table}
\centering
\footnotesize
\caption{Comparison of LRM and CRM against LES results.}
\vspace{2pt}
\label{tab:MR_model_comparison_improved}
\setlength{\tabcolsep}{2.5pt}
\renewcommand{\arraystretch}{1.1}
\vspace{2pt}
\begin{tabular}{c ccc cc ccc cc}
\multicolumn{11}{l}{\textit{Rectangular constrictions}} \\[-6pt]
\hline
& \multicolumn{5}{c}{$M_I=1.4$} & \multicolumn{5}{c}{$M_I=1.8$} \\
\cline{2-6} \cline{7-11}
\noalign{\vskip 4pt}
BR & LES & LRM & $\Delta \text{LRM}$\% & CRM & $\Delta \text{CRM}$\% 
& LES & LRM & $\Delta \text{LRM}$\%  & CRM & $\Delta \text{CRM}$\% \\[-4pt]
\hline
0.30 & 1.109 & 1.096 & -1.17 & 1.099 & -0.92 & 1.293 & 1.267 & -2.01 & 1.316 & +1.75 \\
0.40 & 1.143 & 1.131 & -1.05 & 1.134 & -0.76 & 1.345 & 1.329 & -1.19 & 1.359 & +1.05 \\
0.50 & 1.176 & 1.166 & -0.85 & 1.169 & -0.56 & 1.396 & 1.391 & -0.36 & 1.402 & +0.42 \\
0.60 & 1.209 & 1.201 & -0.66 & 1.205 & -0.36 & 1.448 & 1.453 & +0.35 & 1.445 & -0.24 \\
0.70 & 1.243 & 1.236 & -0.56 & 1.240 & -0.23 & 1.499 & 1.515 & +1.07 & 1.488 & -0.76 \\
0.75 & 1.260 & 1.254 & -0.48 & 1.258 & -0.14 & 1.525 & 1.547 & +1.44 & 1.509 & -1.03 \\
\hline
\noalign{\vskip -4pt}
\multicolumn{1}{l}{RMS } & & & 0.83 & & 0.57 & & & 1.22 & & 1.0 \\
\noalign{\vskip -4pt}
\hline
\end{tabular}
\vspace{6pt}

\begin{tabular}{c ccc cc ccc cc}
\multicolumn{11}{l}{\textit{Sinusoidal constrictions}} \\[-6pt]
\hline
& \multicolumn{5}{c}{$M_I=1.4$} & \multicolumn{5}{c}{$M_I=1.8$} \\
\cline{2-6} \cline{7-11}
\noalign{\vskip 4pt}
BR & LES & LRM & $\Delta \text{LRM}$\% & CRM & $\Delta \text{CRM}$\% 
& LES & LRM & $\Delta \text{LRM}$\%  & CRM & $\Delta \text{CRM}$\% \\[-4pt]
\hline
0.30 & 1.059 & 1.096 & +3.46 & 1.049 & -0.92 & 1.248 & 1.267 & +1.49 & 1.255 & +0.55 \\
0.40 & 1.099 & 1.131 & +2.89 & 1.093 & -0.54 & 1.302 & 1.329 & +2.06 & 1.309 & +0.52 \\
0.50 & 1.138 & 1.166 & +2.45 & 1.136 & -0.22 & 1.356 & 1.391 & +2.58 & 1.361 & +0.34 \\
0.60 & 1.177 & 1.201 & +2.05 & 1.177 & +0.04 & 1.410 & 1.453 & +3.06 & 1.412 & +0.12 \\
0.70 & 1.217 & 1.236 & +1.59 & 1.220 & +0.21 & 1.464 & 1.515 & +3.51 & 1.463 & +0.09 \\
0.75 & 1.236 & 1.254 & +1.45 & 1.241 & +0.39 & 1.491 & 1.547 & +3.72 & 1.488 & +0.18 \\
\hline
\noalign{\vskip -4pt}
\multicolumn{1}{l}{RMS } & & & 2.42 & & 0.48 & & & 2.85 & & 0.35 \\
\noalign{\vskip -4pt}
\hline
\end{tabular}
\end{table}
The validity of representing all losses with a single constant rests on two conditions: that separation, rather than distributed viscous dissipation, is the dominant loss mechanism, and that $C_c$ is insensitive to the shock strength driving the flow. Both are satisfied within the moderate range $M_I = 1.4$ and $1.8$ examined here. Sensitivity analysis shows that perturbations of $\pm 0.10$ to either calibrated value raise the RMS error by at most $1.6$ percentage points and keep all errors below $2.7\%$. At significantly higher incident Mach numbers, the post-shock temperature rise and resulting changes in boundary-layer displacement thickness may introduce a Mach-dependent correction to $C_c$; validating this regime would require dedicated experiments or simulations beyond the scope of the present study. The predictions of CRM are given in Table \ref {tab:MR_model_comparison_improved} and are compared with the LES results. The disparity between CRM and LES is calculated as $\Delta\text{CRM} \% =(\text{CRM}-\text{LES})/\text{LES}\times100$. The choked-flow model achieves RMS errors of $0.57\%$ \& $1.00\%$ for the rectangular cases and $0.48\%$ \& $0.35\%$ for the sinusoidal cases, reducing errors by roughly $30\%$ for rectangular cases and $80$-$90\%$ for sinusoidal cases relative to LRM; the improvement is larger for sinusoidal geometries because the slope in the  LRM is tied to $M_{RW}$, which overestimates the compression for a smooth-contour geometry, whereas the introduction of $C_c$ captures the geometry effects.

The choked-flow model recovers the correct physics at both extremes of the blockage ratio, providing an independent consistency check. At the upper limit $\mathrm{BR} \to 1$, the effective blockage $\mathrm{BR}_\mathrm{eff} \to 1$, and the area ratio $A_c/A^*_\mathrm{eff}$ in equation~(5.8) diverges. Satisfying the choking 
constraint then requires $M_2 \to 0$, which from \eqref{eq:state1}-\eqref{eq:state2} corresponds to $u_2 = 0$: 
The post-reflected flow is brought to rest, recovering the classical rigid-wall reflection condition $M_R \to M_{RW}$ exactly. Numerically, the choked-flow model yields $M_R = 1.3515$ at $\mathrm{BR} = 0.999$ and $M_I = 1.4$, against the exact wall-reflection value $M_{RW} = 1.3519$; the agreement at $M_I = 1.8$ is equally close ($1.6216$ versus $1.6221$). The linear model shares this upper limit by construction (equation \eqref{eq:MRW}), so both models are consistent with the wall-reflection bound originally expressed in equation \eqref{eq:MRW} of the main text. At the lower limit $\mathrm{BR} \to 0$, $\mathrm{BR}_\mathrm{eff} \to 0$ and the required area ratio in equation~\eqref{eq:MR_closure} approaches unity. In this case the throat imposes no additional contraction, the choking constraint exerts no compression on the upstream flow, and therefore $M_R \to 1$: no reflected shock is generated in the absence of blockage. 

\subsection{Predictive model for the transmitted shock wave}

The evolution of the transmitted shock wave is governed by a fundamentally different set of physical mechanisms than those controlling the reflected shock wave. Whereas the reflected shock wave forms through an almost instantaneous reflection process, the transmitted disturbance propagates through a region in which the flow undergoes strong lateral expansion, separation, and subsequent relaxation through multiple reflections before re-establishing a quasi-one-dimensional structure downstream (see figure \ref{ref_schem}).

A useful physical framework for predicting this behaviour was established previously by \citet{gichon2024dynamics}, who investigated the propagation of shock waves far downstream of abrupt area expansions. There, it was shown that once a shock diffracts into an expanded region, it does not immediately recover a uniform normal structure. Instead, the shock undergoes a prolonged relaxation process governed by repeated lateral reflections, which collectively redistribute momentum across the conduit cross-section. As a result, the shock strength far downstream can be related to the expansion ratio through a relaxation-type decay law rather than an instantaneous area-change relation. In its reduced-order form, the downstream transmitted Mach number following an expansion was expressed as a gradual relaxation from the inlet value, with the decay rate controlled by the effective expansion ratio and the associated shock relaxation timescale. This formulation captures the cumulative weakening induced by lateral shock spreading and repeated wall interactions as the flow evolves towards a pseudo-steady normal shock.

The present configuration shares key physical similarities with the expansion problem studied by \citet{gichon2024dynamics}; however, two important differences must be taken into account. First, the shock transmitted through the constriction does not propagate in a purely expanding environment. Although the core of the wave experiences lateral expansion upon exiting the constricted region, it is simultaneously subjected to secondary compression induced by reflected shocks from the interaction with the contraction section that trails the primary front. Second, the transmitted shock inherits additional streamwise momentum from the accelerated flow upstream of the constriction. The pre-compression and flow acceleration preceding the constriction effectively increase the impulse carried by the disturbance as it enters the expanded section. Consequently, the downstream evolution reflects a competition between expansion-driven relaxation and reflection-induced strengthening. To capture this combined behaviour, we adopt the relaxation-type framework proposed by \citet{gichon2024dynamics} as a baseline description for the expansion-driven decay, and introduce corrective terms that account for (i) partial shock reinforcement by trailing reflected waves from the constriction inlet and (ii) the augmented momentum flux imposed by the upstream compression process.

To account for the fact that the transmitted shock does not immediately expand into the full conduit height upon exiting the constriction, we introduce an effective conduit hight $H_c$, which represents the lateral extent of the emerging jet and is bounded above by the unobstructed conduit height $H$ as

\begin{equation}
    H_c = {H} \left(\min\!\left(1,\, \alpha(1-\mathrm{BR})\right)\right)
\end{equation}

The coefficient $\alpha$ accounts for the augmented streamwise momentum carried by the flow as it enters the constriction from the wider upstream conduit. Effectively, $\alpha$ acts similarly to $C_c$ in the constriction entrance where it accounts for effects of added momentum which in this case reduces the effective area to which the flow expands thus increasing the momentum  downstream of the constriction.  
Using $H_c$ we can define an effective expansion ratio
\begin{equation}
    \mathcal{E}_{\mathrm{eff}} =\frac{H}{H_C}= \frac{1}{\min\!\left(1,\,\alpha(1 - \mathrm{BR})\right)}
\end{equation}

Guided by gradual-expansion shock-decay behavior,  we can formulate a transmitted shock wave model, TM, that predicts the far downstream transmitted shock wave Mach number as

\begin{equation}
M_T \approx 1 + \beta\,\mathcal{E}_{\mathrm{eff}}^{\,n}(M_I-1)
\end{equation}

\begin{table}
  \centering
  \caption{Predicted transmitted shock wave Mach number from the reduced-order model.}
  \label{tab:MT_model_comparison}
  \small
  \setlength{\tabcolsep}{6pt}
  \renewcommand{\arraystretch}{1.1}
\vspace{8pt}

\begin{tabular}{@{}c ccc ccc@{}}
  \multicolumn{7}{l}{\textit{Rectangular constrictions}} \\
    \toprule
    & \multicolumn{3}{c}{$M_I = 1.4$} & \multicolumn{3}{c}{$M_I = 1.8$} \\
    \cline{2-4}\cline{5-7}
    \noalign{\vskip 4pt}
    BR & LES & TM & $\Delta \text{TM}$\% & LES & TM & $\Delta \text{TM}$\% \\
    \midrule
    0.30 & 1.367  & 1.3868 & $-$1.45 & 1.738 & 1.7736 & $-$2.05 \\
    0.40 & 1.357  & 1.3736 & $-$1.22 & 1.712 & 1.7471 & $-$2.05 \\
    0.50 & 1.338  & 1.3423 & $-$0.32 & 1.670 & 1.6845 & $-$0.87 \\
    0.60 & 1.310  & 1.3075 & $+$0.19 & 1.612 & 1.6150 & $-$0.19 \\
    0.70 & 1.273  & 1.2678 & $+$0.41 & 1.539 & 1.5357 & $+$0.21 \\
    0.75 & 1.227  & 1.2454 & $-$1.50 & 1.450 & 1.4908 & $-$2.81 \\
    \midrule
    \multicolumn{1}{@{}l}{RMS } & & & 1.01 & & & 1.69 \\
    \bottomrule
  \end{tabular}

 \vspace{10pt}

  \begin{tabular}{@{}c ccc ccc@{}}
    \multicolumn{7}{l}{\textit{Sinusoidal constrictions}} \\
    \toprule
    & \multicolumn{3}{c}{$M_I = 1.4$} & \multicolumn{3}{c}{$M_I = 1.8$} \\
    \cline{2-4}\cline{5-7}
    \noalign{\vskip 4pt}
    BR & LES & TM & $\Delta \text{TM}$\% & LES & TM & $\Delta \text{TM}$\% \\
    \midrule
    0.30 & 1.381  & 1.387  & $+$0.43 & 1.739 & 1.774  & $+$2.01 \\
    0.40 & 1.370  & 1.375  & $+$0.36 & 1.722 & 1.751  & $+$1.68 \\
    0.50 & 1.350  & 1.348  & $-$0.15 & 1.687 & 1.695  & $+$0.47 \\
    0.60 & 1.321  & 1.317  & $-$0.30 & 1.634 & 1.633  & $-$0.06 \\
    0.70 & 1.283  & 1.281  & $-$0.16 & 1.563 & 1.561  & $-$0.13 \\
    0.75 & 1.2614 & 1.260  & $+$0.11 & 1.521 & 1.5195 & $-$0.10 \\
    \midrule
    \multicolumn{1}{@{}l}{RMS } & & & 0.28 & & & 1.14 \\
    \bottomrule
    \noalign{\vskip 4pt}
  \end{tabular}
\end{table}

The exponent $n$ reflects the relaxation rate of the expanding compression wave as it adjusts to the downstream conduit and is taken from \citet{gichon2024dynamics}, yielding $n = -0.48$ for a step expansion and $n = -0.42$ for a gradual expansion, thereby linking the relaxation behaviour to the internal structure of the constriction. The coefficients $\beta = 0.967$ and $\alpha = 1.55$ are empirical constants obtained from a fit to all of the LES sets. 

The TM predictions are plotted in figure~\ref{M_R_M_T}(\textit{c}) and (\textit{d}), with quantitative comparisons summarized in table~\ref{tab:MT_model_comparison}. The predictions of TM are compared with the LES results. The disparity between TM and LES is calculated as $\Delta \text{TM}\% =( \text{TM}-\text{LES})/\text{LES}\times100$. Overall, the agreement between the reduced-order model and the numerical simulations is very good across all geometries and blockage ratios. The corresponding mean RMS disparities are $1.05\%$ for $M_I = 1.4$ and $1.72\%$ for $M_I = 1.8$ in the rectangular cases and $0.28\%$ for $M_I = 1.4$ and $1.14\%$ for $M_I = 1.8$ . These values indicate that the dominant dependence of the transmitted shock strength on blockage ratio and relaxation dynamics is well captured in all four data sets.

Slightly larger deviations are observed for the most open constrictions at the higher incident Mach number. In these cases, the stronger interaction within the constriction generates secondary reflected structures that subsequently catch up with the transmitted shock during its downstream propagation, leading to deviation from the simple relaxation description. Nevertheless, even in these cases, the discrepancy remains limited, confirming that the present model captures the primary physics governing late-time transmitted shock attenuation while neglecting only higher-order interaction effects.

\section{Conclusions}

This study presents a systematic investigation of shock-wave interaction with a localised constriction in a straight conduit, by varying geometry type, blockage ratio, and constriction length. By considering abrupt rectangular and smoothly contoured sinusoidal constrictions as two well-defined limiting cases, we isolated and interpreted the coupled roles of the constriction geometry in governing both the transient start-up dynamics and the evolution towards a steady reflected and transmitted shock waves.

We have found that, counterintuitively, the constriction length plays fundamentally different roles in the two geometric cases, and does so in opposing directions. For rectangular constrictions, the reflected shock strength is governed almost exclusively by blockage ratio and is insensitive to length. This follows directly from the instantaneous head-on interaction at the upstream vertical face, which establishes the upstream compression before the internal flow has time to adjust. The transmitted shock, by contrast, shows a measurable length dependence as 
longer constrictions improve pressure matching across the constriction and produce a more coherent downstream jet, thereby sustaining a stronger transmitted shock. For sinusoidal constrictions, the role of length is reversed. Because the incident shock encounters a continuously varying wall slope, the reflection topology evolves along the contour and depends on the maximum surface angle. Shorter sinusoidal constrictions, with steeper maximum slopes, promote regular reflection and generate strong reflected shocks comparable to those of rectangular geometries. Longer, more gradual constrictions sustain a Mach reflection configuration over much of the contour, distributing the upstream compression over a larger streamwise distance and substantially weakening the reflected shock. As a result, in sinusoidal geometries, the reflected shock depends sensitively on both blockage ratio and length, while the transmitted shock, which more gradually adjusts to the contracting- expanding geometry, is comparatively insensitive to geometric details. The rectangular constriction thus provides a length-independent upper bound on the strength of the reflected shock for a given blockage ratio.

The transient start-up process within the constriction strongly depends on the geometry and is dynamically rich, involving separation, vortex formation, shear-layer development, and the establishment of standing shock systems. In all cases examined, this adjustment process occurs over time scales approximately two orders of magnitude longer than the passage time of the incident shock through the constriction. This clear separation of time scales establishes that the initial shock-geometry interaction and the subsequent internal flow adjustment are physically distinct processes. We have found that stronger constrictions, which generate stronger reflected shocks, somewhat unexpectedly produce shorter start-up times, as the more energetic wave system drives a faster rearrangement of the internal flow. 

Despite the pronounced differences in transient behaviour across geometries and constriction lengths, the late-time reflected and transmitted shock strengths collapse onto well-defined, regular trends that depend primarily on blockage ratio, with constriction length and geometry type acting as secondary modifiers. This behaviour underscores a key physical insight: although the short-time dynamics differ markedly between cases, the cumulative outcome of these complex interactions converges toward a state dominated by the blocked ratio. The geometry type and length determine the pathway to this state and the rate at which the internal flow adjusts, but they do not fundamentally alter the late-time strength of the reflected and transmitted fronts. Specifically, the reflected shock Mach number scales linearly with blockage ratio across the full parameter space, for the different geometry types, while the transmitted shock follows a monotonic relaxation-type decay with increasing blockage ratio, reflecting the progressive redistribution of energy as the shock expands through the constricted region. These observations provide the physical justification for developing semi-empirical reduced-order models. For the reflected shock, we have developed two simplified models. A Linear shock reflection model (LRM) was proposed based on the assumption of linearity between the two limiting cases of an open and a fully blocked conduit. The second, choked-flow model (CRM) was developed based on isentropic area-Mach relations, with a single calibrated contraction coefficient $C_c$ that accounts for the vena contracta of the flow that goes through the constriction. Both models perform well, with CRM outperforming LRM, achieving RMS disparities below $2\%$ for rectangular geometries and below $3\%$ for sinusoidal geometries relative to the LES results. For the transmitted shock, a relaxation-based model (TM) incorporating an effective expansion ratio that accounts for the augmented streamwise momentum of the emerging jet achieves RMS disparities below $2\%$ for both rectangular and sinusoidal geometries.

Taken together, the results presented here establish a consistent physical picture linking local geometric features to the global evolution of shock systems in confined flows. By resolving both the transient start-up dynamics and the late-time propagation of reflected and transmitted shocks across a broad parameter space, this study clarifies how blockage, length, and contour jointly govern energy partitioning, wave structure, and the resulting impulsive flow field. The semi-empirical models developed herein further translate these physical insights into predictive capability, enabling accurate estimation of shock strengths under realistic conditions. These findings provide a framework that can be directly applied to a wide range of applications involving internal compressible flows with localized geometric variations, including propulsion systems, blast mitigation strategies, and transient flows in complex conduits, where the interplay between geometry and wave dynamics ultimately governs performance, loading, and stability. 
\newpage

\appendix
\section{Minimum blockage ratio to form choked flow}
\label{app:choking}
\setcounter{table}{0}
\renewcommand{\thetable}{\thesection\arabic{table}}

\begin{figure}
   \centerline{\includegraphics[scale=0.8]{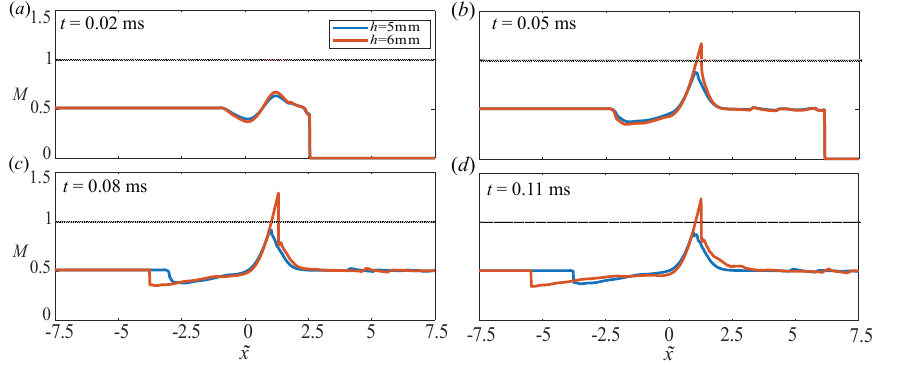}}
   \caption{Mach number distribution of the flow  along the centerline of the computational domain plotted for cases with $\tilde{L}$=2 and $h=5mm$(BR = 0.25) and $h=6mm$(BR = 0.3) at $t=20,50,80,110 \,\mu s$} 
   \label{Mach_for_05_06}
\end{figure}

To establish the range of blockage ratios to examine, we first identify the minimum blockage ratio that causes choking in the throat. We can assume that in a sinusoidal constriction with relatively long $L$ and low $h$ the assumption of isentropic flow holds when its most narrow section is just narrow enough to achieve choked flow. Under this assumption, we can calculate the critical blockage ratio corresponding to achieve choked flow can be determined using the area-Mach number relation \citep{anderson1990modern}:
\begin{equation}
\frac{1}{M}\left[\frac{2} \gamma+1\cdot\left(1+\frac{\gamma-1}{2}M^2\right)\right]^\frac{\gamma+1}{2\left(\gamma-1\right)}=\frac{A}{A^*}
\label{A2A*}
\end{equation}
In our case, the domain is two-dimensionless so ${A}/{A^*}=H/H^*$ 
Applying $\gamma=1.4$ for air, initial mach number of $M=1.4$, and tunnel height of $H$=40 mm$^2$ yields that in order to achieve choked flow, the minimum cross section of the tunnel should be $H^*$=29.86mm. Therefore, for contractions higher than $h\approx 5.5$mm, the flow will should be choked inside the constriction. Fig. \ref{Mach_for_05_06} presents LES results performed with an impinging shock wave with incident Mach number of $M_I$ = 1.4 and $L$ = 80 mm showing the pressure distribution along the centerline of the computational domain for 2 different sinusoidal geometries with $h$ = 5 mm and 6 mm, at different time steps. Figure \ref{Mach_for_05_06} indicates that initially, as the shock wave passes through the constriction, the flow behind it is subsonic. However, the pressure across the constriction quickly builds up until it stabilizes. For the case of \( h = 6 \, \text{mm} \), the local Mach number within the constriction becomes supersonic, meaning that the flow becomes choked. Since this is the least constrictive case, we can assume that any configuration shorter than $L$= 80 mm and narrower than the $h$ = 6 mm (BR=0.3) case will result in choked flow.

\section{Empirical fitting parameters}
\setcounter{table}{0}
\label{appendix_C}

To provide a quantitative representation of the measured trends, the reflected and transmitted shock strengths presented in figure \ref{M_R_M_T} are approximated by simple functions of BR. The reflected shock Mach number \(M_R\) is represented by a linear function, while the transmitted shock Mach number \(M_T\) is represented by a quadratic function,
\begin{equation}
M_T = a (BR)^2 + b (BR) + c,
\qquad
M_R = d (BR) + e,
\label{eq:empirical_fits}
\end{equation}

These functional forms are chosen to reflect the approximately linear dependence of \(M_R\) and the nonlinear dependence of \(M_T\) over the measured parameter range. The coefficients are obtained by least-squares regression applied to the data shown in figure~\ref{M_R_M_T}. The resulting parameters for the two incident Mach numbers considered are summarised in table~\ref{tab:fit_parameters}.

\begin{table}
\centering
\caption{Fitting parameters obtained by least-squares fitting of function \eqref{eq:empirical_fits} to reflected and transmitted shock wave Mach numbers in figure \ref{M_R_M_T}.}
\label{tab:fit_parameters}
\setlength{\tabcolsep}{6pt}
\renewcommand{\arraystretch}{1.0}

\begin{tabular}{lcccc}
\toprule
 & \multicolumn{2}{c}{Sinusoidal} & \multicolumn{2}{c}{Rectangular} \\
\cmidrule(lr){2-3} \cmidrule(lr){4-5}
Parameter & $M_I = 1.4$ & $M_I = 1.8$ & $M_I = 1.4$ & $M_I = 1.8$ \\
\midrule

$a$ & $-1.1\times10^{-3}$ & $-2.2\times10^{-3}$ & $-2\times10^{-3}$ &$-1.1\times10^{-3}$  \\

$b$ & $9.8\times10^{-3}$  & $2.2\times10^{-2}$  &  $15\times10^{-3}$  & $1.5\times10^{-2}$  \\

$c$ & $1.36$              & $1.69$              &  $1.72$              & $1.35$  \\

$d$ & $2.0\times10^{-2}$  & $2.7\times10^{-2}$  &  $2.6\times10^{-2}$  & $1.7\times10^{-2}$  \\

$e$ & $9.4\times10^{-1}$  & $1.09\times10^{0}$  &  $1.14\times10^{0}$  & $1.01\times10^{0}$  \\

\bottomrule
\end{tabular}
\end{table}

\section{LES versus geometrical shock dynamics comparison}
\setcounter{table}{0}
\label{app:GSDLES}
As introduced in Section \ref{Large eddy simulations}, the results presented in this work are obtained from large-eddy simulations (LES) of the compressible Navier-Stokes equations. As a reduced alternative for tracking shock-front propagation, geometrical shock dynamics (GSD) relates the local shock strength to geometric variations of the shock front while neglecting post-shock flow effects \citep{whitham1957new,whitham1959new, henshaw1986numerical}. GSD has been extended and applied to converging shocks and shocks in varying area conduits, including shock propagation through converging-diverging channels \citep{ndebele2019interaction}, shock strengthening in smooth contractions and subsequent attenuation in expansions \citep{whitham1958propagation,bird1959effect,russell1967shock,dowse2014area}, and converging polygonal shocks \citep{apazidis2002experimental}. In the present configuration, GSD is used to estimate the streamwise evolution of the transmitted-shock strength through the constriction. The goal of this appendix is to assess the accuracy of this approximation by direct comparison with the corresponding LES predictions.

In GSD, the evolution of the shock Mach number is governed by the area-Mach number relation
\begin{equation}
    A = A_0 \frac{f(M)}{f(M_0)},
\end{equation}
where $A$ and $M$ are the local ray-tube area and shock Mach number, and $A_0$ and $M_0$ denote their values at a reference location. The function $f(M)$ is defined by
\begin{equation}
    f(M) = \exp \left( - \int \frac{M \lambda(M)}{M^2 - 1} \, dM \right),
\end{equation}
with
\begin{equation}
    \lambda(M) = \left(1 + \frac{2}{\gamma+1} \frac{1-\mu^2}{\mu} \right)
    \left(1 + 2\mu + \frac{1}{M^2} \right), 
    \quad
    \mu^2 = \frac{(\gamma-1)M^2 + 2}{2\gamma M^2 - (\gamma-1)}.
\end{equation}
Given $A/A_0$, the relation above yields $M$ implicitly and therefore a prediction for the spatial variation of transmitted-shock strength that can be compared directly with LES. The specific implementation and solution strategy are further discussed in \citet{gichon2024dynamics}. 

\begin{figure}
   \centerline{\includegraphics[scale=0.85]{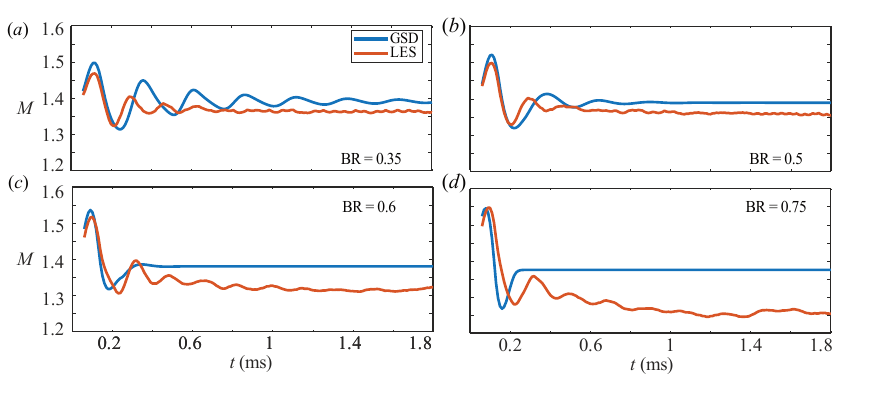}}
   \caption{Mach number of transmitted shock wave obtained from GSD (blue) and LES(orange), for sinusoidal geometries with $\tilde{L}$=2 and BR =  (a) 0.35 (b) 0.5 (c) 0.6 (d) 0.75} 
   \label{GSD_VS_LES_SHOCK_LOCATION}
\end{figure}

Figure \ref{GSD_VS_LES_SHOCK_LOCATION} presents a comparison of GSD and LES results of the evolution of the transmitted shock wave Mach number in time, following the interaction of an $M_I$ = 1.4 shock wave with a constriction of length $\tilde{L}=2$ and BR= 0.35,0.5,0.6,0.75. The results show large disparities between GSD and LES, which is unsurprising, since GSD assumes free propagation independent of post-shock and reflected-flow effects. In a constricted geometry, strong wall-induced disturbances and reflected waves interact with the shock, modifying both its strength and structure. In addition, choking within the constriction imposes a mass flow limitation that significantly weakens the transmitted shock, an effect that is not captured by GSD.

 \begin{acknowledgments}
 \end{acknowledgments}

\bibliographystyle{jfm} 
\bibliography{newbib}
\end{document}